\documentclass{article}
\pdfoutput=1 % if your are submitting a pdflatex (i.e. if you have
\usepackage{jheppub} % for details on the use of the package, please
\usepackage[utf8]{inputenc}
\usepackage{amsmath}
\usepackage{amssymb}
\usepackage{mathrsfs}
\usepackage{amsfonts}
\usepackage{amsthm}
\usepackage{appendix}
\usepackage{graphicx,hyperref,bbm}
\usepackage{soul}
\usepackage{subfloat}

\def\twind{w}
\def\bx{\chi^0}
\def\by{\chi^1}
\def\SL2R{\textit{SL}$(2,\mathbb{R})$}

 \author{Yasuaki Hikida$^1$}
 \author{Volker Schomerus$^{2,3}$}
\affiliation{$^1$ Center for Gravitational Physics and Quantum Information, Yukawa Institute for Theoretical Physics, Kyoto University, Kitashirakawa Oiwakecho, Sakyo-ku, Kyoto 606-8502, Japan}
\affiliation{$^2$ DESY Theory Group, Deutsches Elektronen Synchroton DESY, Notkestrasse 85, 22603 Hamburg, Germany}
\affiliation{$^3$ II. Institut f\"ur Theoretische Physik, Universit\"at Hamburg, Luruper Chaussee 149, D-22761 Hamburg}
\affiliation{Zentrum f\"ur Mathematische Physik, Universit\"at Hamburg, Bundesstrasse 55, D-20146 Hamburg }

\emailAdd{yhikida@yukawa.kyoto-u.ac.jp}
\emailAdd{volker.schomerus@desy.de}

\title{Engineering Perturbative String Duals\\ for Symmetric Product Orbifold CFTs}
\date{December 2023}

\abstract{Constructing a holographic string theory dual for a CFT in the perturbative, weakly coupled
regime is a holy grail for gauge/string dualities that would not only open the door for proofs
of the AdS/CFT correspondence but could also provide novel examples of string duals with and without
supersymmetry. In this work we consider some marginal perturbation of a family of symmetric
product orbifolds in two dimensions. From their correlation functions we engineer a bosonic string
theory whose amplitudes are shown to reproduce the CFT correlation function order-by-order both in
the coupling and in $1/N$. Our derivation does not require to compute and compare correlation functions
explicitly but rather relies on a sequence of identities that can be derived using path integral
methods. The bosonic string theory we engineer is based on the field content of the Kac-Wakimoto
representation of strings in $AdS_3$ with $k$ units of pure NSNS flux, but the interaction terms
we obtain are different. They include current algebra preserving interaction terms
with one unit of spectral flow.}

\arxivnumber{2312.05317}
\preprint{YITP-23-159}

\begin{document}
 \addtolength{\baselineskip}{2mm}
\maketitle

\section{Introduction}

About 25 years after Maldacena proposed the first concrete example of a 't Hooft-like gauge-string
duality \cite{Maldacena:1997re}, the AdS/CFT correspondence has become the key source for novel insight
into non-perturbative Quantum Field Theory (QFT) on the one hand and into quantum gravity on the other.
In the early years, supergravity constructions were exploited to assemble an extensive zoo of such
dualities in various dimensions and with various amounts of supersymmetry. Note that the existence of
some geometric supergravity regime on the string theory side was build into the search strategy
from the start. This was not seen as a significant drawback but rather became the most celebrated
feature of the AdS/CFT correspondence: It allowed to compute quantities in strongly coupled field
theories through the geometry of the dual supergravity. On the other hand, strong/weak coupling
dualities are of course notoriously difficult to verify.

In the decade after the first examples of the AdS/CFT correspondence had been proposed, testing its
implications becames a major focus of the field. This started with protected quantities that could be
calculated exactly with the help of supersymmetry and were shown to interpolate successfully between
the supergravity regime and the weakly coupled field theory. The famous circular Wilson loop 
\cite{Drukker:2000rr} in $\mathcal{N}=4$ Superymmetric Yang-Mills (SYM) theory provides a prototypical 
example. With the rise of integrability based techniques that were initiated by Minahan and Zarembo 
\cite{Minahan:2002ve}, precision tests of unprotected quantities also became available. The most 
prominent example of this kind was uncovered in the context of the cusp anomalous dimension of 
${\mathcal N}=4$ SYM theory. In a seminal paper Beisert-Eden-Staudacher \cite{Beisert:2006ez} proposed 
a remarkably simple integral equation that could be used to compute the cusp anomalous dimension for 
any value of the coupling. In particular, the so-called BES equation gives access to both weak and
strong coupling expansions and the latter were checked to very high precision against gauge
theory calculations on the one hand and string calculations on the other. The form of the BES
equation resembles the integral equations that are familiar from the Thermodynamic Bethe Ansatz
(TBA) in 1+1-dimensional integrable quantum field theory. While the appearance of such an
equation seems surprising from the gauge theory perspective, it is rather natural from the
dual side. After all, the Gubser-Klebanov-Polyakov string \cite{Gubser:2002tv} that captures
the deformation of the cusp anomalous dimension away from the supergravity regime is a
1+1-dimensional quantum system. For a gauge theory practitioner, on the other hand, the BES
equation remains truly remarkable since it miraculously manages to repackage gauge theory
calculations in such a way that even perturbative precision calculations can be performed on
a sheet of paper or on a laptop.

Moving on from extensive tests to genuine proofs of the AdS/CFT correspondence has been a holy
grail of the field ever since Maldacena proposed the first example. While this still seems out
of reach for higher dimensional theories, in spite of very advanced and powerful techniques,
there has been some progress in lower dimensions recently. The simplest non-trivial examples
of the AdS/CFT correspondence appear for 1-dimensional systems of quantum mechanics. In this
case the best studied examples involve Jackiw-Teitelboom (JT) gravity or deformations thereof
on the dual side. In particular, the holographic relation between a new critical string theory, 
dubbed the Virasoro minimal string, and certain double scaled matrix integrals was recently 
proven in \cite{Collier:2023cyw}. This includes the famous duality between JT gravity and 
matrix integrals of Saad et.\ al.\ \cite{Saad:2019lba} in a certain limit. Moving up in 
dimensions to 2-dimensional conformal field theories a few holographic dualities are on  
equally solid ground. In particular, it was shown in \cite{Eberhardt:2019ywk,Eberhardt:2020akk,
Dei:2020zui} that the worldsheet of superstring on $AdS_3 \times S^3 \times T^4$ with the minimal 
unit of NSNS flux localizes to particular Riemann surfaces, which essentially derives the 
AdS$_3$/CFT$_2$ duality between the minimal tension superstrings and the orbifold point of 
symmetric product orbifold $T^4/S_N$.

One would certainly hope that proofs of the AdS/CFT correspondence can eventually lead to
derivations which engineer new examples of dual string theories directly from the perturbative
formulation of the gauge theory. As we have reviewed above, the BES equation remains valid all
the way to vanishing 't Hooft coupling in the gauge theory. In this sense, even the free gauge
theory admits a stringy description. While the historical path has lead us to think of the
gauge theory in terms of strings on $AdS_5 \times S^5$, it seems likely that there exists another
dual description of the string theory that captures the perturbative expansion around the point
where the curvature radius $R$ of $AdS_5$ assumes a minimal value. If such a string dual indeed
exists, it would provide a string theoretic dual of type IIB theory on $AdS_5 \times S^5$,
somewhat akin to the string theoretic dual one can construct at the Gepner points of
Calabi-Yau compactifications, see \cite{Witten:1993yc} and references therein. This
nurtures hopes that one day we might actually be able to turn the perturbative formulation
of gauge theory into the worldsheet formulation of some (non-geometric) string theory and
then use the technology of 1+1-dimensional (integrable) models to perform gauge theory
calculations with the astounding efficiency of BES-like equation. We refer to this kind of
repackaging of gauge theory degrees of freedom as \textit{engineering} of a dual string
theory. Note that such an engineering process could extend the usual AdS/CFT correspondence
since it does require the existence of a large volume regime in which the string theory
becomes geometric.

While this may still seem like a far fetched vision for higher dimensional gauge theories,
it now seems to come within reach, at least in lower dimensions. Indeed, what we will lay
out below may be considered as the first concrete example of engineering a perturbative
string dual in the sense we described. The context of our discussion is an interacting 
2-dimensional CFT that can be obtained from special symmetric product orbifolds with
parent theory $\mathcal{M}$ by switching on a particular marginal interaction. The duality
of this theory with strings on some $AdS_3$ compactification with pure NSNS flux was first
proposed and tested in a very remarkable paper by Eberhardt \cite{Eberhardt:2021vsx}, 
see also \cite{Argurio:2000tb,Balthazar:2021xeh,Martinec:2021vpk} for some related works. 
Here we shall take this duality significantly further by rewriting the complete set
of orbifold correlators at any order in the coupling constant and in $1/N$ as a
bosonic string theory. After some significant massaging of the orbifold correlators
using techniques we had developed in \cite{Hikida:2007tq,Hikida:2008pe}, we will be 
able to read off the field content and the interactions of the dual string background 
without ever computing a single correlator explicitly.

\begin{figure}
\begin{subfigures}
 \begin{minipage}[t]{0.25\textwidth}
  \begin{center}
   \includegraphics[keepaspectratio, width=27mm]{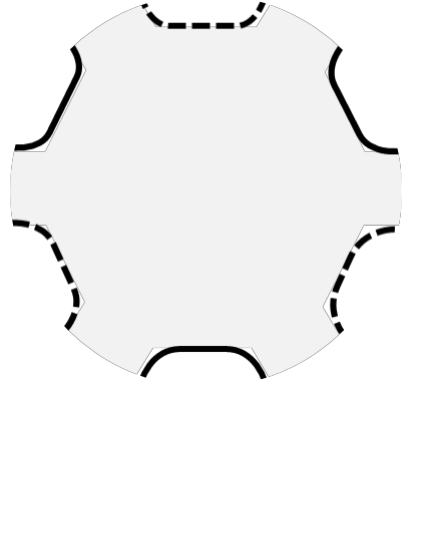}\\[1cm]
  \end{center}
  \vspace*{-6mm} 
  \caption{Graphical representation of a twist field with cycle length $w=3$.}
  \label{fig:1a}
 \end{minipage}
  \hspace{0.03\textwidth}
 \begin{minipage}[t]{0.40\textwidth}
 \begin{center}
  \includegraphics[keepaspectratio, width=50mm]{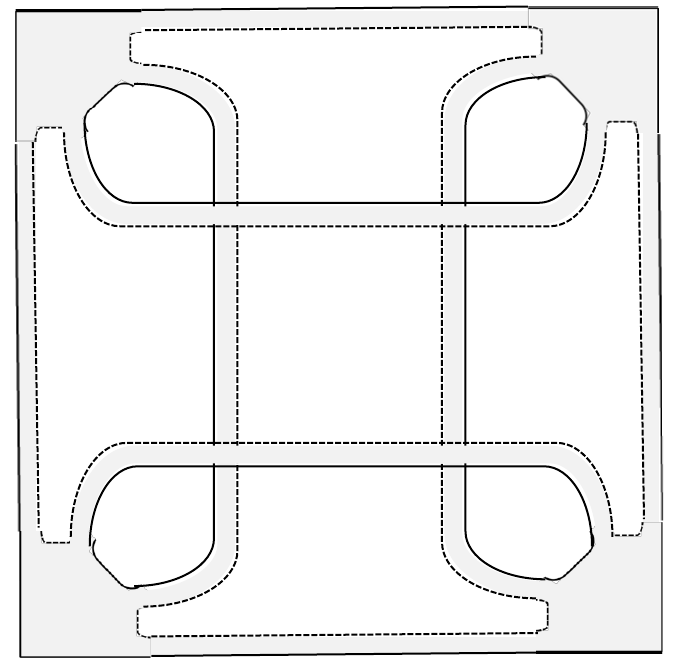}
 \end{center}
  \caption{Feynman diagram for a 4-point function of fields with $w=2$ drawn on a surface of $g=1$ with 2 solid loops (and same number of dashed ones).}
  \label{fig:1b}
 \end{minipage}
\end{subfigures}
\hspace{0.03\textwidth}
 \begin{minipage}[t]{0.26\textwidth}
 \begin{center}
  \includegraphics[keepaspectratio, width=23mm]{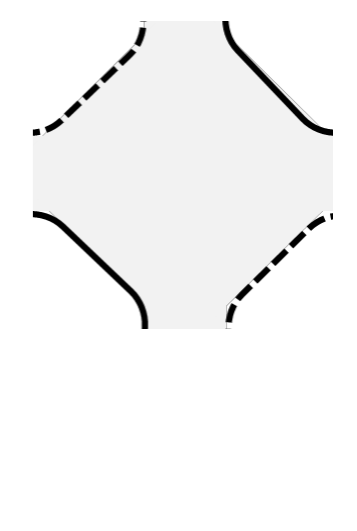}
 \end{center}
  \caption{Graphical repre\-sentation of marginal 
operator in the twisted sector with $w=2$.}
  \label{fig:2}
 \end{minipage}
\end{figure}

In order to describe our main constructions and results in more detail, we will begin
with a brief review of the relevant symmetric product orbifolds. Regardless of the
specific parent theory $\mathcal{M}$ one uses, symmetric product orbifolds possess a
diagrammatic representation that resembles the diagrammatic representation of gauge
theories. In both cases, propagators are represented as double lines. But unlike
ordinary gauge theories, the two lines representing propagators in symmetric product
orbifolds need to be distinguished. Following \cite{Pakman:2009zz} we will denote one
by a solid and the other by a dashed line. The analogue of single trace operators in
gauge theory are fields in single cycle twist fields of the symmetric product orbifold.
The length $w$ of the cycle controls the length of the operator, i.e. the number of Wick
contractions it is involved in. We represent such operators as shown in Figure \ref{fig:1a}
through
a circle with $2w$ emanating propagator lines. The Feynman graphs that can be drawn by
connecting the double lines at the vertices with double line propagators must satisfy two
conditions that were first described in \cite{Pakman:2009zz}. The precise formulation
of these constraints is not relevant for our discussion. One example of a diagram
that represents one of the contributions to the 4-point function of twist fields
with $w_\nu=2, \nu = 1, \dots, 4$ is depicted in Figure  \ref{fig:1b}. As one can easily see,
it can be drawn on a surface of genus $g=1$. More generally, the genus $g$ of the
surface is related to the length parameters $w_\nu, \nu=1, \dots, n$ of $n$ fields
and the number $R$ of closed solid loops through
\begin{equation} \label{eq:gRrel}
2-2g  = 2R - \sum_\nu w_\nu   + n \ .  
\end{equation} 
Eberhardt duality proposal involves a particular marginal operator in the $w=2$ twisted
sector of the symmetric orbifold theory \cite{Eberhardt:2021vsx}, see also
\cite{Balthazar:2021xeh}. A concrete formula will be given below. The
pictorial representation of this operator is given by a 4-point vertex, see Figure \ref{fig:2}.
In the perturbed symmetric orbifold theory, this operators is introduced with a
parameter $\lambda$ that controls the strength of the coupling.

There is a well explored way to encode various diagrams that contribute to a
correlation function in the symmetric product orbifold through covering maps. These
covering maps $\Gamma: \Sigma_g \mapsto S^2$ describe an $R$-fold branched covering
of the sphere on which we evaluate the orbifold correlators by a surface of genus $g$.
The branching numbers of the branched cover are determined by the cycle lengths
of the fields in the correlator.
\smallskip

We shall propose some precise (non-geometric) bosonic string theory and show that it
is indeed dual to perturbed symmetric product orbifold. The string theory reproduces
the CFT correlators we described in the previous two paragraphs order-by-order in the
coupling constant $\mu$. The underlying worldsheet CFT with $c=26$ is a product of
some unitary CFT $\mathcal{X}$ with some appropriate Wess-Zumino-Novikov-Witten (WZNW)
model. The field content of the latter is given by a bosonic $\beta\gamma$ system along
with a linear dilaton $\phi$ with background charge $Q_\phi = b = 1/\sqrt{k-2}$,
\begin{equation} \label{eq:S0}
S_0[\beta,\gamma,\varphi] = \frac{1}{2\pi}\int d^2 z
\left(\partial \phi \bar \partial \phi -  \beta\bar \partial \gamma - \bar \beta
\partial \bar \gamma + \frac{Q_\phi}{4} \sqrt{g} \mathcal{R} \phi\right) \ .
\end{equation}
To this free field theory we add two marginal interaction terms. The first one is given by
\begin{equation}\label{eq:Qm}
S_\text{int}^-[\beta,\gamma,\varphi] =  
- \int d^2v \gamma(v)^{1-k} \bar\gamma(\bar v)^{1-k} v^{(-1)}(v, \bar v) \,
e^{\frac{1}{b}\phi(v,\bar v)}
\end{equation}
where $v^{(-1)}$ denotes some (spectrally flowed) vacuum state of the $\beta\gamma$
system, see below. The interaction term $S^-_\text{int}$ is a very close relative of the ``screening
charge of second kind'' that was found by Bershadski and Ooguri in \cite{Bershadsky:1989mf}. 
Insertions of the interaction vertex \eqref{eq:Qm} are needed in order to reproduce the free 
orbifold theory. In fact, the number of insertions coincides with the number of sheets in 
the branched covering of $S^2$.\footnote{While we were making final edits to our paper 
we learned that the same interaction term also appears in some upcoming works by 
Dei et al.\ and Knighton et al.\ \cite{Dei:2023ivl,Knighton:2023mhq}.  See also 
\cite{Giveon:2019gfk} for an earlier incarnation in a different context.} 

In order to also reproduce the interaction terms from the marginal perturbation of the 
orbifold theory one has to introduce a second interaction term in the worldsheet string 
theory. The latter is given by
\begin{equation} \label{eq:Qp}
S_\text{int}^+[\beta,\gamma,\varphi]  =  - \lambda \int d^2u \gamma(u) \bar \gamma(\bar u) 
v^{(1)}(u,\bar u) \ .
\end{equation}
This interaction term has not appeared in discussion of the string dual to symmetric
product orbifolds before, see however \cite{Giveon:2019gfk} for a different context. Within 
the correlation functions we consider, it possesses some unusual properties. In particular, 
it produces a trivial contribution when it is inserted in maximally winding number 
violating correlation functions as long as it stays away from other insertion points. It 
turns out, however, that non-trivial contributions of the interaction term \eqref{eq:Qp} 
can arise. They stem from regions in the moduli space of field insertions in which pairs 
of these vertex operators collide. Once such a pair is formed it is able to reproduce the 
insertion of the interaction vertex in the dual symmetric product orbifold. Let us stress 
that the theory \eqref{eq:S0} along with the two interaction terms \eqref{eq:Qm} and 
\eqref{eq:Qp} differs from the sigma model on $AdS_3$, even though the latter can be 
expressed through the same field content. The relation between our worldsheet theory and 
the usual worldsheet model for strings in $AdS_3$ is rather reminiscent of the relation 
between Liouville field theory and its nonperturbative dual that is obtained by sending 
$b$ to $1/b$. This is of course not entirely unexpected. 
\smallskip 

Our construction of this worldsheet model, including the two screening charges we
have displayed above is entirely systematic. Its derivation rests on two pillars. The 
first pillar is a novel embedding of the symmetric product orbifolds into the a free 
worldsheet theory with Kac-Wakimoto field content. This embedding is realized concretely 
in eqs.\ \eqref{corrSL2R} and \eqref{corrSL2Rg} for tree level and higher genus 
contributions, respectively. These two formulas represent the central new results of 
this work.\footnote{Some avatars have already appeared in \cite{Hikida:2020kil}. The
analysis we shall present in our work below elaborates and improves on the previous 
study. In addition, we interpret the resulting formulas more widely in the context 
of Eberhardt's holographic duality.}
The relation is quite reminiscent of the relation between Liouville field 
theory and the $H_3^+$ WZNW model \cite{Ribault:2005wp}. In our setup, the role of 
Liouville field theory is played by the perturbed symmetric product orbifold whose 
interaction is indeed of Liouville form, though dressed with some twist field. And 
indeed, the tools we use in order to proof our formulas are mostly borrowed from our 
path integral derivation of the Liouville-$H^+_3$ duality, see \cite{Hikida:2007tq}, 
with some extensions from \cite{Hikida:2008pe} and some new ingredients. 

At this point we have just rewritten correlations functions of twist fields in the 
symmetric product orbifolds through correlation functions of vertex operators with 
non-trivial spectral flow in the Kac-Wakimoto free field theory. But in comparison to 
the conventional rewriting in terms of higher genus correlators of the parent theory, 
see above, all the complicated prefactors of the latter get completely absorbed 
through our way of rewriting the correlations functions in terms of Kac-Wakimoto 
vertex operators. While conceptually, our embedding just provides an equality 
between sets of correlators in two different CFTs, the remarkable absorption of 
prefactors is decisive for our ability to uplift the correlation functions 
to string theory. This uplift exploits the second important pillar of the 
construction, namely the observation that the string theory contains a field 
$\gamma(z)$ which localizes to a covering map $\Gamma: \Sigma_g \mapsto S^2$ 
in computations of string amplitudes. The map $\Gamma$ turns out to coincide 
with the covering map that perturbative expansion of the symmetric product 
orbifold, see above. The localization of the field $\gamma$ and its relation 
with the branching functions $\Gamma$ go back to works of Eberhardt, Gaberdiel 
and Gopakumar, see e.g. \cite{Eberhardt:2019ywk} and references therein.
\smallskip 

Let us finish this introduction with a brief outline. In the next section we will
give a rather self-contained introduction into those symmetric product orbifolds
that participate Eberhardt's duality. In particular we shall state a precise
formula that expresses correlation functions of the orbifold theory in terms of
correlators of the parent theory on branched coverings of the sphere. A particular
focus is on the properties of the associated covering maps $\Gamma$. Section \ref{section.3}
contains some background material on \SL2R WZNW models and their free
field realization. We will also review the importance of spectrally flowed sectors
and construct the relevant vertex operators. At the end of section \ref{section.3} we have
thereby collected all the material that is needed to establish a new realization of
orbifold correlators in terms of correlation functions of the WZNW model. Our important 
new formula is stated and proved in section \ref{section.4}, using techniques from
earlier work on the $H_3^+$-Liouville relation \cite{Ribault:2005wp} and the closely 
related Fateev-Zamolodchikov-Zamolodchikov (FZZ) duality \cite{FZZ}, see 
\cite{Hikida:2007tq,Hikida:2008pe}. This result is then applied
in section \ref{section.5} to argue that the string amplitudes of the bosonic string 
theory described through eqs. \eqref{eq:S0}, \eqref{eq:Qm} and \eqref{eq:Qp} indeed
reproduce the correlation functions of the orbifold theory. The arguments
are based on the localization mechanism for the field $\gamma$ that we briefly 
sketched in the previous paragraph. In section \ref{section.6} we extend all the 
above beyond the leading terms of the planar theory by considering coverings 
$\Gamma$ associated with Riemann surfaces $\Sigma_g$ of genus $g > 0$, see also 
\cite{Eberhardt:2020akk}. This work concludes  with an extensive list of 
interesting extensions as well as applications.

\section{Symmetric Product Orbifolds and Their Correlations}
\label{section.2}

The starting point of this work is provided by a certain class of symmetric product orbifolds.
Here we shall provide a short self-contained introduction on their field content and correlation
functions. With a focus on the particular parent theory that appears in the context of holography
we will construct and discuss the fields that are dual to one-particle states on the string theory
side of the correspondence. These include one particular marginal operator that we will use to
deform the symmetric product orbifolds. Correlation functions in symmetric product orbifolds
can be related to correlators of the parent theory through so-called covering maps. This will
be reviewed in the second subsection.

\subsection{Symmetric product orbifolds and their perturbation}

Symmetric product orbifolds of the form $\mathcal{M}^N/S_N$ can be associated with
any 2-dimensional conformal field theory $\mathcal{M}$. We refer to the latter as the
\textit{parent theory} an denote its central charge by $c_\mathcal{M}$. The central
charge of the associated symmetric product orbifold is given by $c = N c_\mathcal{M}$.

Even though much of what we shall review below is largely independent of the particular
choice of the parent CFT $\mathcal{M}$ we shall introduce the relevant one right away.
Following \cite{Eberhardt:2019qcl} we consider products $\mathcal{M} = \mathcal{X}
\otimes \mathbb{R}_\varphi$ of some unitary conformal field theory $\mathcal{X}$
and a (free) linear dilaton $\varphi$. This means that the state space of the
parent theory splits as
\begin{equation} \label{eq:parentM}
\mathscr{H}_\mathcal{M}  = \mathscr{H}_\mathcal{X} \otimes \mathscr{H}_\varphi\ .
\end{equation}
Here, the first factor $\mathscr{H}_\mathcal{X}$ is spanned by the states $\psi$
of the CFT $\mathcal{X}$. We shall denote the left- and right-moving conformal weights
of the states $\psi$ by $h_\psi$ and $\bar h_\psi$, respectively. The central charge
$c_\mathcal{X}$ of the CFT $\mathcal{X}$ is supposed to be of the form
\begin{equation} \label{eq:cX}
 c_\mathcal{X} = 26 - \frac{3k}{k-2}\
\end{equation}
with some parameter $k$. In order to motivate this parametrization we anticipate that
the worldsheet theory of the dual string background will turn out to be a product of
the CFT $\mathcal{X}$ and an \SL2R WZNW model at level $k$. The latter has central
charge $c_k = 3k/(k-2)$. Hence, eq.\ \eqref{eq:cX} will eventually ensure that the
worldsheet CFT gives rise to a consistent critical bosonic string theory.

The second factor in the product on the right hand side of equation \eqref{eq:parentM}
refers to the state space of a linear dilaton CFT, i.e. the free conformal field theory
for a single bosonic field $\varphi$ with background charge
\begin{equation} \label{eq:Qvarphi}
 Q=Q_\varphi = b - \frac{1}{b} \quad \textit{where} \quad b^{-2} = k-2 \ .
\end{equation}
Note that the background charge of the linear dilaton is tied to the parameter $k$ that
we introduced in eq.\ \eqref{eq:cX}. In our conventions, the choice of background charge
implies that the associated central charge of the linear dilaton theory is given by
$$ c_\varphi = 1+ 6 Q_\varphi^2 = 1 + \frac{6(k-3)^2}{k-2}\ .$$
Following standard conventions, the vertex operators of the linear dilaton CFT will be
denoted by $V_\alpha$ and their left- and right-moving weights by $h_\alpha, \bar h_\alpha$,
with
\begin{equation} \label{eq:Valpha}
V_\alpha = e^{2\alpha\varphi} \quad \textit{and} \quad
h_\alpha = \bar h_\alpha = \alpha (Q_\varphi - \alpha) \ .
\end{equation}
To be quite precise, the usual normal ordering prescriptions must be applied in order for
these vertex operators to be well defined in the quantum theory. Such a prescription is
simply assumed throughout the entire text.

Putting the two factors together, we obtain the parent CFT $\mathcal{M}$ with combined
central charge given by
$$ c_\mathcal{M} = c_\mathcal{X} + c_\varphi =  26 - \frac{3k}{k-2} +
     1 + \frac{6(k-3)^2}{k-2}= 6k $$
and states/fields of the form
\begin{equation} \label{eq:Valphapsi}
V_\alpha(\psi) := \psi \, e^{2 \alpha \varphi} \quad \textit{ with }
\quad h_\alpha(\psi) = h_\psi + \alpha (Q_\varphi-\alpha) \ , \
\end{equation}
and a similar expression for $\bar h_\alpha(\psi)$ with $h_\psi$ replaced by
$\bar h_\psi$. This concludes our brief description of the relevant parent CFT
$\mathcal{M}$ and we can now turn to the associated symmetric product orbifold.
\medskip

In order to construct the symmetric product orbifold $\textit{Sym}^N(\mathcal{M}) =
\mathcal{M}^N /S_N$ we prepare $N$ of identical copies of the parent CFT $\mathcal{M}$
and then perform an orbifold construction with respect to the action of the symmetric
group $S_N$ that permuted the $N$ identical copies. According to the standard rules,
the resulting orbifold conformal field theory has twisted sectors that are labeled by
the conjugacy classes of the symmetric group. Let $\omega \in S_N$ be any permutation
of $N$ objects. We shall denote by $[\omega]$ the associated conjugacy class, i.e. the
set of elements that are related to $\omega$ by conjugation with any other element,
by $[\omega]$. As is well known, any permutation can be decomposed into a product of
single cycle permutations. In the dual string theory, twisted sector states in a
single cycle permutation are represented by single particle states. Therefore, our
discussion will focus on the twisted sectors that are associated with conjugacy
classes of single cycle permutations. There is one such conjugacy class $[w]$ for
each cycle length $w \leq N$. We shall denote the ground states of the associated
twisted sectors by $\mathcal{O}_w = \mathcal{O}_{[w]}(p)$.%
\footnote{ In order to make the twist operator to be invariant under the action of symmetric group, we need to sum over all its images. Properly normalizing the two-point function, $n$-point function is proportional to $N^{1 - g - n/2}$ for large $N$, which leads to the relation to the string coupling $g_s^2 \sim 1/N$, see \cite{Jevicki:1998bm,Lunin:2000yv,Pakman:2009zz}.}
According to standard
results form the theory of orbifolds, these ground states have conformal weight
\begin{align}
 h_w = \bar h_w =
 \frac{c_\mathcal{M}}{24} \left ( w - \frac{1}{w} \right)
  = \frac{k}{4}\left ( w - \frac{1}{w} \right)\, .
\end{align}
By acting with the fields \eqref{eq:Valphapsi} from the parent theory $\mathcal{M}$
we can obtain the following set of twisted sector states/fields
\begin{equation} \label{eq:Vwalphapsi}
V^{(w)}_\alpha(\psi) :=
|w|^{\frac{k}{2} \frac{(w - 1)^2}{w}}  \psi \, e^{2 \alpha \varphi}\,
\mathcal{O}_w \ ,
\end{equation}
for all $w \leq N, \alpha \in Q_\varphi/2 + i \mathbb{R}$ and $\psi \in
\mathscr{H}_\mathcal{X}$. Here we assumed the twist fields $\mathcal{O}_w$ to be
canonically normalized and included a $w$-dependent prefactor by hand that will
turn out to be convenient later on. The conformal weights of these fields are
given by
\begin{equation} \label{eq:hwalphapsi}
h^{(w)}_\alpha(\psi) = \frac{ h_\psi + \alpha (Q_\varphi - \alpha)}{w} +
\frac{k}{4} \left( w - \frac{1}{w} \right) \ ,
\end{equation}
and a similar expression for the weight $\bar h^{(w)}_\alpha(\psi)$.
Let us stress that the first term is obtained from the conformal weights in the
parent theory, see eq.\ \eqref{eq:Valphapsi}, by division with the cycle length
$w$. This reflects the fact that in the twisted sector the boundary conditions
are changed so that the field only comes back to itself after circling
$w$ times around the twist field $\mathcal{O}_w$.
\smallskip

We are now finally prepared to identify the marginal operator we want to deform the
symmetric orbifold theory by, see  \cite{Eberhardt:2021vsx}, and to define the set of
correlation functions we will analyse throughout the remainder of this work. Given the
formulas we displayed in the previous paragraph it is easy to verify that the
following operator
\begin{align} \label{marginalop}
V_{\frac{1}{2b}}^{(2)} = 2^{ \frac{k}{4}}  V_{\frac{1}{2b}}\, \mathcal{O}_{2}\
\end{align}
is indeed marginal.
Let us stress that we have chosen the state $\psi$ of the CFT $\mathcal{X}$ to be given
by the vacuum state $\psi = |0\rangle$. With the parameter $\alpha = 1/2b$ and the cycle
length $w=2$, our formula \eqref{eq:hwalphapsi} indeed evaluates to $h^{(2)}_{1/2b}=1$.
In the process of this calculation one needs to insert the expression \eqref{eq:Qvarphi}
for the screening charge of the linear dilaton theory. The operator \eqref{marginalop}
resembles the interaction term of Liouville field theory. Indeed, it involves a single 
exponential of the linear dilaton field $\varphi$ which is now dressed with the twist 
field $\mathcal{O}_2$. As in Liouville field theory, one can expect the resulting model 
to be conformal. {Let us note that there exists a second `dual' marginal 
operator that from a reflection $\alpha \rightarrow Q_\varphi - \alpha$ of the momentum 
in the linear dilaton theory, 
\begin{align} \label{dmarginalop}
V_{b-\frac{3}{2b}}^{(2)} = 2^{ \frac{k}{4}}  V_{b-\frac{3}{2b}}\, \mathcal{O}_{2}\ . 
\end{align}
This marginal operator should not be added to the action simultaneously with the 
interaction \eqref{marginalop}. But it provides a second screening charge that 
explains additional poles in the correlation functions of the interacting model 
and can be used to calculate their residues through free field calculations.}

Up to now we have avoided to specify the insertion point of our fields. Throughout this
work we will deal with correlation functions in which the fields of the symmetric product
orbifold are inserted on a sphere. Insertion points in the sphere will be denoted by the
letters $x,y$, to distinguish them for insertion points on higher genus surfaces that
will enter our discussion soon. The subject of our interest are $n$-point functions of
fields in the perturbed symmetric product orbifold on the sphere,\footnote{Throughout 
this work we assume all correlation functions to be normalized, i.e. we always divide 
by the corresponding partition functions.}
\begin{align} \label{corrdef}
 \langle \prod_{\nu=1}^n V^{(w_\nu)}_{\alpha_\nu} (\psi_\nu;x_\nu)
 \rangle^{S^2}_\lambda := \sum_{s=1}^\infty \frac{\lambda^s}{s!}  \, \langle \prod_{\nu=1}^n
 V^{(w_\nu)}_{\alpha_\nu} (\psi_\nu;x_\nu) \ \int \prod_{a=1}^{s}
 d ^2 y_a V^{(2)}_{\frac{1}{2b}} (y_a)  \, \rangle_0^{S^2} \, ,
\end{align}
where $\psi_\nu \in \mathscr{H}_\mathcal{X}$ and the subscript $0$ indicates that the
correlation function in the integrand on the right hand side is evaluated by the undeformed
symmetric product orbifold. We note that for correlation functions in the linear dilaton
theory to be non-zero, the fields must satisfy the charge condition
\begin{align} \label{eq:alphaconstraint}
 \sum_{\nu=1}^n \alpha_\nu + \frac{s}{2b} = Q_\varphi \ .
\end{align}
This is entirely analogous to Liouville field theory, where correlation functions depend
meromorphically on the parameters $\alpha_\nu$ and develop poles whenever these parameters
satisfy charge conditions. It is the residues of these poles that can be computed in
perturbation theory. 
{We note that there are additional poles. They arise from insertions of 
the dual screening charge that is associated with the marginal operator \eqref{dmarginalop}. 
The presence of $r$ such insertions modifies the charge condition \eqref{eq:alphaconstraint} 
by adding a term of the form $r(b-3/2b)$ on the left hand side. In order not to expand formulas 
below we shall not carry these dual screening through our analysis. Since it is in principle 
straight forward to include the second screening charge it will suffice to add some comments 
from time to time.}

\subsection{Correlations functions and covering maps}

It has been appreciated for a long time that correlation functions of twist fields in
orbifold theories on a sphere can be evaluated in terms of correlation functions of the
parent theory on some appropriate branched cover. We will first review the general
construction in the case of symmetric product orbifolds (see, e.g. \cite{Lunin:2000yv,
Lunin:2001pw,Pakman:2009zz}) before discussing some
specific aspects that concern the specific correlations functions \eqref{corrdef}
that arise in the context of our perturbative deformation.

So, let us consider some set of fields $V_{\alpha_A}^{(w_A)}(\psi_A;x_A)$ of the form
\eqref{eq:Vwalphapsi} with the label $A$ running through $A=1, \dots, m = n+s$. Later
we will set $w_A = 2, \alpha_A = 1/2b$ and $\psi_A = |0\rangle$ for $A > n$, but
for the moment that is irrelevant. As stated in the previous section, all these
fields are inserted on a Riemann sphere, i.e.\ a surface of genus $g=0$. It turns
out that the correlation function of these fields can be evaluated by considering
correlation functions of the associated fields $V_{\alpha_A}(\psi_A;z_A)$ in the
parent theory. The latter, however, must now be inserted at the branching points
of an $R$-sheeted branched covering over the sphere. The order of the $m$ branch
points is given by the integers $w_A$ that specify the relevant twist fields in
the orbifold theory. According to the Riemann-Hurwitz formula, the number $R$ of
sheets, the ramification numbers $w_A$ and the genus $g$ of the covering surface
$\Sigma_g$ are constrained by the equation\footnote{There is a slight subtlety 
here: Because of the prefactor 1/2 in front of the sum over ramification numbers, 
the resulting $R^{(g)}$ need not be integer. Whenever that happens, $R^{(g)}$ 
cannot be interpreted as a number of sheets. For ramification numbers that lead 
to half-integer $R^{(g)}$ the corresponding correlation function vanishes.}
\begin{equation} \label{eq:Rg}
R^{(g)} = \frac12 \sum_{A=1}^{m} (w_A - 1) + 1 - g \ .
\end{equation}
For the time being we want to focus on the contributions from covering surfaces
of genus $g=0$, leaving the extension to coverings of higher genus to section \ref{section.6}.

The relevant $R$-sheeted branched coverings of the Riemann sphere by a surface
$\Sigma_0$ of genus $g=0$ can be nicely encoded in a meromorphic covering map
$\Gamma: \Sigma_0 \rightarrow S^2$ with the following behavior
\begin{align}
 \Gamma(z) = x_A + a^\Gamma_\nu (z - z_A)^{w_A} + \mathcal{O}
          ((z - z_A)^{w_A+1})
 \label{Gammadef}
\end{align}
near the branch points at $z = z_A$ of the surface $\Sigma_0$. The constant
term $x_A$ in the expansion is determined by the insertion point of the
twisted sector field in the orbifold correlator.

The form \eqref{Gammadef} of the local expansions near the branch points
imposes very strong constraints on $\Gamma$ which eliminate any continuous
parameters. This is best analysed by looking at the derivative $\partial
\Gamma$. The expansions \eqref{Gammadef} obviously imply that $\partial
\Gamma$ has a zero of order $w_A-1$ at $z=z_A$. Hence, the total order of
zeroes is given by $2R^{(0)} - 2$. Since $\partial \Gamma$ is meromorphic
of weight one, we conclude that it must have $R = R^{(0)}$ second order
poles, i.e. it must be of the form
\begin{align} \label{partialGamma}
    \partial \Gamma (z) = \frac{q \prod^{m}_{A=1} (z - z_A)^{w_A -1}}
    {\prod_{j =1}^R (z - v_j)^2} \ .
\end{align}
This expression contains a total number of $m+R+1$ parameters, namely the
positions $z_A$ of zeroes, the positions $v_j$ of poles and the constant
prefactor $q$. Upon passing from $\partial \Gamma$ to $\Gamma$ we gain one
more integration constant. Now let us look at the constraints. The total
residue of $\partial \Gamma$ vanishes by construction, but since $\partial
\Gamma$ is a total derivative, all the local residues at $v_j$ must vanish.
This imposes $R-1$ extra conditions. After integration, we also need to
impose the relations $\Gamma(z_A) = x_A$ which are clearly not built into
eq.\ \eqref{partialGamma}. This gives $m$ conditions and leaves us with
just $3$ parameters. The latter are of course associated with the global
conformal symmetry of the orbifold correlation function we consider. These 
may be fixed by requiring that $z_1 = 0, z_2 = 1$ and $z_3 = \infty$. From 
now on we shall assume this choice. We have now shown what we anticipated, 
namely that the covering map $\Gamma$ is fully determined by the insertion 
points $x_A$, up to some remaining discrete choices that are associated with 
different Feynman graphs. This also means that the positions $z_A$ of the 
branch points, the positions $v_j$ of the poles, the prefactor $q$ as well 
as the coefficients $a^\Gamma_A$ in eq.\ \eqref{Gammadef} should be considered 
as functions of the insertion points $x_B$, i.e.
\begin{equation}
z_A = z_A^\Gamma = z_A(x_B) \ , \quad
v_j = v_j^\Gamma = v_j(x_B) \ , \quad q = q^\Gamma = q(x_B) \ , 
\quad a_A^\Gamma = a_A(x_B)\ .
\end{equation}
In some simple cases the dependence can be calculated explicitly, but in
general this poses a very challenging problem one cannot hope to solve,
especially for larger number of insertion points in the orbifold theory.

Given the notion of the covering map it is now possible to give a precise
formula for the relation between the correlation function of the symmetric
product orbifold on the sphere and the correlation function of the parent
theory on the covering surface $\Sigma_0$. With a proper regularization of
divergences, see e.g. \cite{Lunin:2000yv,Hikida:2020kil}, the final result is
\begin{align} \label{freecorr}
\begin{aligned}
& \langle \prod_{A=1}^{m} V^{(w_A)}_{\alpha_A}(\psi_A;x_A) \rangle^{S^2} = \\ 
& \quad =
 \sum_{\Gamma} \prod_{A =1}^{m} (\tilde a_A^\Gamma)^{- h^{(w_A)}_{\alpha_A}(\psi_A)
 + \frac{k}{4} (w_A -1)}
 (\bar{\tilde a}_A^\Gamma)^{- \bar h^{(w_A)}_{\alpha_A}(\psi_A)
 + \frac{k}{4} (w_A -1)}
 \prod_{j =1}^R |\xi^\Gamma_j|^{- k} \langle  \prod_{A=1}^{m}
 V_{\alpha_A}(\psi_A;z_A) \rangle^{\Sigma_0} \, .
 \end{aligned}
\end{align}
Here the sum runs over the discrete set of covering maps. For the leading planar
contributions, these covering maps satisfy eq.\ \eqref{Gammadef}. The exponents in
the summands contain the weights that were defined previously in eq.\ \eqref{eq:hwalphapsi}
and we have introduced functions
\begin{align} \label{axi}
\tilde a_A^\Gamma = w_A a^\Gamma_A
= \frac{q \prod_{B \neq A} (z_A - z_B)^{w_B -1}}{\prod_{j=1}^R (z_A - v_j)^2} \, , \quad
\xi^\Gamma_j = \frac{q \prod_{A=1}^{m} (v_j - z_A)^{w_A -1}}{\prod_{i \neq j} (v_j - v_{i})^2} \, .
\end{align}
Note that the positions of zeroes $z_A$ and poles $v_j$ of $\partial \Gamma$ are
functions of the insertion points $x_B$ on the left hand side, as we explained
above. We also note that the left hand side is the leading contribution to the
correlator on the right hand side in the $1/N$ expansion. Correction arise from
branched coverings of higher genus.
\smallskip

We conclude this subsection with a few specific comments on the integrated correlation
functions \eqref{corrdef} we are going to analyse below. To apply the formulas \eqref{freecorr}
and \eqref{axi}, we set $m = n+s$ and set $w_A=2, \alpha_A = 1/2b, \psi_A = |0\rangle$ and
$x_A = y_{A-n}$ for all $A > n$. By plugging eq.\ \eqref{freecorr} into eq.\ \eqref{corrdef}
we obtain
\begin{align} \label{correv}
 &\langle \prod_{\nu=1}^n V^{(w_\nu)}_{\alpha_\nu} (\psi_\nu;x_\nu)
 \rangle^{S^2}_\lambda  := \sum_{s=1}^\infty \sum_\Gamma \frac{\lambda^s}{s!}
 \int \prod_{a=1}^{s}\left(
 d ^2 y_a |\tilde b^\Gamma_a|^{-2} \right) \times \\
 & \times 
 \prod_{\mu=1}^{n } 
(\tilde a_\mu^\Gamma)^{-h_{\alpha_\mu}^{(w_\mu)} + \frac{k}{4} (w_\mu -1)} (\bar{\tilde a}_\mu^\Gamma)^{-\bar h_{\alpha_\mu}^{(w_\mu)} + \frac{k}{4} (w_\mu -1)} \prod_{a=1}^s |\tilde b^\Gamma_a|^{\frac{k}{2}} \prod_{j=1}^{R_0+s/2}  |\xi_j^\Gamma|^{-k}
 \langle \prod_{\nu=1}^n
 V_{\alpha_\nu} (\psi_\nu;z_\nu)  \prod_{a=1}^{s} V_{\frac{1}{2b}} (u_a)  \rangle^{\Sigma_0}\, , \nonumber
\end{align}
where we introduces $\tilde b_a^\Gamma = \tilde a_{n+a}^\Gamma$ for $a=1, \dots, s$.
Let us note that the number $R$ of poles of the relevant covering map $\Gamma$ does depend on
the number $s$ of insertion of the marginal operator. Indeed, according to formula \eqref{eq:Rg}
we find
$$ R^{(0)} = \frac12 \sum_{A=1}^{n+s} (w_A-1) + 1 = \frac12 \sum_{\nu=1}^n (w_\nu-1) +
\frac{s}{2} +1 \equiv R^{(0)}_0 + \frac{s}{2}\ .$$
Of course, the number of poles must be an integer. As was discussed in some detail also in
\cite{Eberhardt:2021vsx}, this condition is only satisfied for $s$ even in case the quantity
$R_0$ is integer.\footnote{As we have pointed out in a footnote for eq.\ \eqref{eq:Rg}, the 
number $R_0$ can also come out to be half-integer. In those cases only odd values of $s$ 
appear in the expansion \eqref{correv} (and the term without insertions of the interaction 
vanishes).} 

The expression we displayed still contains an integration over the insertion points $y_a$
of the perturbing operator of the orbifold model. The integrand, on the other hand, does no
longer depend on the $y_a$ explicitly, but only implicitly through the locations $z_\nu$ and
$u_a$ of the branch points on $\Sigma_0$ as well as through the functions $\tilde a^\Gamma_\nu,
\tilde b^\Gamma_a$ and $\xi_j$. It is certainly natural to change the integration variables from
the parameters $y_a$ on the original sphere $S^2$ to the parameters $u_a$ on the covering
surface $\Sigma_0$, i.e. to the insertion points of the fields $V_{1/2b}$ on the right hand
side.  After this change of variables we have
\begin{align} \label{deformedcorr}
 \begin{aligned}
 \langle \prod_{\nu=1}^n V^{(w_\nu)}_{\alpha_\nu} (\psi_\nu;x_\nu)
 \rangle^{S^2}_\lambda & := \sum_{s=1}^\infty \sum_\Gamma\frac{\lambda^s}{s!}
 \int \prod_{a=1}^{s}
 d^2 u_a  \prod_{\mu=1}^n
(\tilde a_\mu^\Gamma)^{-h_{\alpha_\mu}^{(w_\mu)} + \frac{k}{4} (w_\mu -1)} (\bar{\tilde a}_\mu^\Gamma)^{-\bar h_{\alpha_\mu}^{(w_\mu)} + \frac{k}{4} (w_\mu -1)} \times \\[2mm]
  & \quad \quad \quad \quad \quad \times  \prod_{a=1}^s |\tilde b^\Gamma_a|^{\frac{k}{2}}\prod_{j=1}^{R_0+s/2}  |\xi_j^\Gamma|^{-k}
 \langle \prod_{\nu=1}^n
 V_{\alpha_\nu} (\psi_\nu;z_\nu)  \prod_{a=1}^{s} V_{\frac{1}{2b}} (u_a)  \rangle^{\Sigma_0}\, .
\end{aligned}
\end{align}
Here we used that the Jacobian for the change of variables is such that $d^2y_a = |\tilde b^\Gamma_a|^2
d^2 u_a$. We stress that this formula looks simpler than it is. In particular we stress that all the
branch points $z_\nu$ were functions of the $y_a$ and hence are now functions of $u_a$, i.e. while we
integrate over the insertion points of $V_{1/2b}$, the other vertex operators are also moved around.
In addition, the various prefactors contain a complicated dependence on the data of the orbifold
correlator. For later use, let us spell these factors out explicitly, see eq.\ \eqref{axi},
\begin{align} \label{abxi}
\begin{aligned}
& \tilde a_\nu^\Gamma = w_\nu a_\nu^\Gamma =   \frac{q \prod_{\mu \neq \nu} (z_\nu - z_{\mu})^{w_{\mu} -1}
\prod_a (z_\nu -u_a)}{ \prod_{j=1}^{R_0+s/2} (z_\nu - v_j)^2 } \, , \\[2mm]
& \tilde b^\Gamma_a =2 b^\Gamma_a
 = \frac{q \prod_{\nu} (u_a - z_{\nu})^{w_\nu -1} \prod_{b \neq a}
 (u_a -u_b)}{ \prod_{j=1}^{R_0+s/2} (u_a - v_j)^2 }
 \, , \\[2mm]
&\xi_j^\Gamma  = \frac{q \prod_\nu (v_j - z_\nu)^{w_\nu -1}\prod_a (v_j - u_a)}{\prod_{i \neq j}(v_j - v_i)^2} \, .
\end{aligned}
\end{align}
While we have now indeed rewritten original correlators of the perturbed orbifold theory  in terms
of integrated correlation functions of the parent theory, the relation is very hard to use for explicit
calculations, especially for higher number of field insertions or higher orders of perturbation theory,
simply because explicit formulas for the covering maps are very hard to come by. In this sense, the
calculation of orbifold correlations remains a rather difficult problem. Fortunately, our rewriting in
terms of a dual string theory will not require such an explicit solution.

\section{Some Background on \textit{SL}\texorpdfstring{$(2,\mathbb{R})$}{(2,\mathbb{R})} WZNW Models}
\label{section.3}

Studies of the WZNW model with non-compact target group
\SL2R have a long history, see in particular \cite{Maldacena:2000hw,Maldacena:2000kv,
Maldacena:2001km} and references to earlier papers therein. In the so-called Kac-Wakimoto
representation their field content is given by a linear dilaton field $\phi$ and a bosonic
$\beta\gamma$ system. Here we shall briefly review how these give rise to representations
of affine Kac-Moody algebras, including those that are obtained by action of the spectral
flow automorphism. Such representations are known from \cite{Maldacena:2000hw,Maldacena:2000kv,
Maldacena:2001km} to play a crucial role in holography. We will also construct vertex operators,
including those carrying non-vanishing spectral flow index, in terms of the Kac-Wakimoto fields.
The second subsection contains a technical result that collects all spectral flow into a single
insertion point by exploiting a relationship with non-compact parafermions. This will play an
important role in our subsequent analysis of orbifold correlators, see section \ref{section.4}.

\subsection{Free field realization of the \textit{SL}\texorpdfstring{$(2,\mathbb{R})$}{(2,\mathbb{R})}
WZNW model}
\label{section.3.1}

\SL2R WZNW models are known to possess a very useful first order formulation in
terms of a scalar field $\phi$ with background charge $Q_\phi$ and a bosonic $\beta\gamma$
system of central charge $c_{\beta\gamma} = 2$. The action of this system is given by
\begin{equation}\label{sl2action}
S [\phi , \beta , \gamma] = S_0 [\phi , \beta , \gamma] + S_\text{int} [\phi , \beta , \gamma]
\end{equation}
where the first non-interacting term reads (see eq. \eqref{eq:S0})
\begin{equation} \label{eq:KWfree}
S_0 [\phi , \beta , \gamma] = \frac{1}{2 \pi} \int d ^2 z \left( \partial \phi \bar \partial \phi -
\beta \bar \partial \gamma - \bar \beta \partial \bar \gamma + \frac{Q_\phi}{4} \sqrt{g}
\mathcal{R} \phi  \right) \ .
\end{equation}
Here the curvature $\mathcal{R}$ is computed from the worldsheet metric  $ds^2 = |\rho(z)|^2 dz
d \bar z$ as
$$ \sqrt{g} \mathcal{R} = - 4 \partial \bar \partial \ln |\rho(z)|^2. $$
We choose the background charge $Q_\phi$ of the linear dilaton conformal field theory to
be given by
$$Q_\phi = b = \frac{1}{\sqrt{k-2}}\ , $$
where $b$ is the same parameter we used to parametrize the background charge $Q_\varphi$ of
the linear dilaton $\varphi$ of the parent theory $\mathcal{M}$ in the previous section.
Note, however, that the two fields have different background charge. The value we choose
for $Q_\phi$ here ensures that the central charge $c_k$ of our model \eqref{eq:KWfree} is
given by
$$ c_k = c_\phi + c_{\beta\gamma} = 1 + 6 Q_\phi^2 + 2 = \frac{3k}{k-2} \ . $$
Hence, once we add the conformal field theory $\mathcal{X}$, the total central charge
assumes the critical value that is necessary for the associated bosonic string theory
to be consistent.

For the time being we are entirely agnostic about the form of the interaction terms except
that we require it to be marginal and to commute with the current algebra symmetry of the
free theory. The latter is famously given by
\begin{align}
\begin{aligned} \label{eq:currents}
&J^+ (z) = - \beta (z) \, , \quad J^3 (z) = - \beta \gamma (z) + b^{-1}
\partial \phi (z) \, , \\[2mm]
&J^- (z) = - \beta \gamma \gamma (z)
+ 2 b^{-1} \gamma \partial \phi (z) - k \partial \gamma (z) \, .
\end{aligned}
\end{align}
There are two well known examples of such interaction terms, but as we shall see later,
the duality with the orbifold model makes a very particular and rather unconventional choice.

\paragraph{Spectral flow and winding number.}
\label{section.2.2}
According to \cite{Maldacena:2000hw} the so-called spectral flow automorphism of the
$\mathfrak{sl}(2)$ current algebra plays a key role in holography. The spectral flow
automorphism is easy to spell out. On the modes of the three currents it acts according
to
\begin{align}
\label{eq:SFonJ}
\rho^w (J_n^3) = J^3_n - \frac{k}{2} w \delta_{n,0} \, ,
\quad \rho^w (J^\pm_n) = J^\pm_{n \pm w} \, .
\end{align}
The automorphism can be used to construct new spectrally flowed representations of
the current algebra from highest weight representations by concatenation. These
contain spectrally flowed vacuum states $|w\rangle$ satisfying
\begin{align}
\rho^w (J_n^a) |w \rangle = 0
\end{align}
for $n \geq 0$ and $a = 3 ,\pm$. The spectral flow automprphim actually descends
from an automorphism of the $\beta\gamma$ system and the free scalar field $\phi$.
By definition, it acts as
\begin{equation} \label{eq:SFonbgp}
\rho^w(\beta_n) = \beta_{n+w} \, , \quad \rho^w(\gamma_n) = \gamma_{n-w} \, ,
\quad \rho^w(a_n^\phi) = a_n^\phi - \frac{k}{2}w \delta_{n,0}
\end{equation}
where $a_n^\phi$ denote the Fourier modes of the field $b^{-1} \partial\phi=
\sum a^\phi_n z^{-n-1}$. It is easy to verify that this action induces the action
\eqref{eq:SFonJ} on the currents \eqref{eq:currents}. In terms of the free
fields, we can now characterize the spectrally flowed vacua $|w\rangle$ as
$$|w\rangle = |w\rangle_{(\beta ,\gamma)} \otimes |w\rangle_\phi$$
where the ghost contribution satisfies
\begin{align} \label{vacuum}
\beta_{n+w} |w\rangle_{(\beta,\gamma)} = \gamma_{n-w}
|w\rangle_{(\beta,\gamma)} = 0 \,
, \quad
\end{align}
for $n \geq 0$ and the second factor $|w\rangle_\phi$ is given by
$$|w \rangle_\phi = e^{- \frac{w}{b} \phi} |0\rangle_\phi \ . $$
In the following we shall frequently pass between states and fields of our WZNW
model. In order to do so, we shall often denote the field $\psi(z,\bar z):= [\psi]
(z,\bar z)$ by placing the associated state $\psi$ in rectangular brackets $[\cdot]$.
For the field that is associated to the spectrally flowed vacuum $|w\rangle$ we use
the following shorthand,
\begin{equation} \label{eq:vw}
v^{(w)}(z,\bar z)  :=  [|w\rangle](z,\bar z)  = [|w\rangle_{(\beta,\gamma)}](z,\bar z)
\, e^{- \frac{w}{b}\phi(z,\bar z)} \ .
\end{equation}
The conformal weight of these spectrally flowed vacua is given by
\begin{equation}
h_w = - \frac12(w^2-w) - \frac{w}{2b}\left(b+\frac{w}{2b}\right)
= - \frac{w^2}{4} k \ .
\end{equation}
This concludes our brief review of spectral flow and winding number in the Kac-Wakimoto
free field representation.

\paragraph{Free field realization of vertex operators.}
The standard vertex operators of the \SL2R WZNW model in the so-called
$m$-basis are given by
\begin{align} \label{eq:vertex}
\Phi^j_{m,\bar m} (z, \bar z)  = N^{j}_{m,\bar m} \,
\gamma^{-j -m} (z)\bar \gamma^{-j - \bar m} (\bar z) e^{2b j \phi (z , \bar z)} \, .
\end{align}
Here $j$ is a complex parameter and $m,\bar m$ are real. In the definition of $\Phi$
we have included a nontrivial normalization
$$ N^{j}_{m,\bar m} = \frac{\Gamma(j+\bar m)}{\Gamma(1-j-m)}\ . $$
This will be convenient later on. As was first explained in \cite{Maldacena:2000hw} in
the context of the AdS/CFT correspondence it is not sufficient to consider these standard
operators. Instead, Maldacena and Ooguri argued that the spectrum of string theory on
$AdS_3$ with pure NSNS flux includes additional states that are obtained by
applying the action of the spectral flow automorphisms. In other words, they suggested
to study the following larger class of vertex operators that is labeled by an additional
spectral flow index
$w \in \mathbb{Z}$,
\begin{align} \label{eq:vertexw}
\Phi^{j,w}_{m,\bar m} (z) = N^{j}_{m,\bar m } \,
[\gamma^{-j -m}_{-w}\bar \gamma^{-j - \bar m}_{-w}|w\rangle]
(z , \bar z) e^{2b (j- \frac{w}{2b^2}) \phi (z , \bar z)}   \, .
\end{align}
It is indeed obtained by acting with the spectral flow automorphism $\rho^w$ on the
standard vertex operators. The latter maps the usual vacuum $|0\rangle$ of the theory
to the spectrally flowed vacua $|w\rangle$ and replaces the zero modes $\gamma_0, \bar
\gamma_0$ by their spectral flow image $\gamma_{-w}, \bar \gamma_{-w}$. In addition, it
generates the additional term $-w/2b^2$ in the exponent.

For our analysis below it will be crucial to switch to the $\mu$-basis. In the free field 
representations, the latter are given by%
\footnote{We implicitly analytically continue $\gamma, \beta$ to take complex values. We also set $\bar \gamma, \bar \beta$ to be complex valued and complex conjugate to $\gamma ,\beta$ at the end of computation. In this sense, we are working on strings on Euclidean AdS$_3$, which is  dual to Euclidean CFT$_2$. The parameters $\mu ,\bar \mu$ introduced in \eqref{eq:mubasis} are complex valued and complex conjugate with each other as well. The same argument holds for $x , \bar x$ introduced in \eqref{xop} below.}
\begin{align} \label{eq:mubasis}
\Phi_j (\mu |z) = |\mu|^{2j} e^{\mu \gamma (z) - \bar \mu \bar \gamma (\bar z)} 
e^{2 b j \phi (z, \bar z)} \, .
\end{align}
From these we can recover the vertex operators \eqref{eq:vertex} in $m$-basis through 
the following simple integral transform
\begin{align} \label{eq:mfrommu}
 \Phi^{j}_{m , \bar m} (z)= \int \frac{d \mu^2}{|\mu|^2} 
 \mu^m \bar \mu^{\bar m} \Phi_{j} (\mu|z) \, .
\end{align}
We can extend this relation to operators with non-trivial spectral flow index $w$ as
\begin{align}
 \Phi^{j,w}_{m , \bar m} (z)= \int \frac{d \mu^2}{|\mu|^2} \mu^m \bar \mu^{\bar m} \Phi_{j,w} 
 (\mu|z) \, .
\end{align} 
The operator on the left hand side of this equality was defined independently through 
equation \eqref{eq:vertexw}. 
 
\subsection{Correlators and parafermionic representation}
\label{section.3.2}

Our discussion below will involve various correlation functions of both integrated and 
unintegrated vertex operators in the free Kac-Wakimoto field theory. The standard correlation 
function of our vertex operators \eqref{eq:vertexw} possesses a path integral representation 
of the form 
\begin{equation} \label{eq:correlationmbasis}
\langle \prod_{J=1}^M \Phi^{j_J,w_J}_{m_J,\bar m_J} (z_J) \rangle^\Sigma = 
\int \mathcal{D} \phi \mathcal{D}^2 \beta \mathcal{D}^2 \gamma\, e^{-S_0[\phi,\beta,\gamma]} 
\prod_{J=1}^M \Phi^{j_J,w_J}_{m_J,\bar m_J} (z_J)\ . 
\end{equation} 
Here we have placed a superscript $\Sigma$ on the expectation value to refer to the surface 
on which the local operators are inserted. Alternatively, we can write these correlation 
functions in terms of vertex operators in the $\mu$-basis as 
\begin{equation} \label{eq:correlationmubasis}
\langle \prod_{J=1}^M \Phi^{j_J,w_J}_{m_J,\bar m_J} (z_J) \rangle^\Sigma = 
\int \prod_{J=1}^M \frac{d \mu_J^2}{|\mu_J|^2} \mu_J^m \bar \mu_J^{\bar m} 
\langle \prod_{J=1}^M \Phi^{j_J,w_J}(\mu_J|z_J) \rangle^\Sigma\ . 
\end{equation} 
For the time we shall restrict to surfaces of genus $g=0$, leaving the discussion of $g > 0$
to section \ref{section.6}. 

We want to introduce a second set of correlation functions that are obtained from the 
standard ones through the insertion of a $\delta$-function constraint in the $\mu$-integral 
on the right hand side of the previous equation
\begin{equation} \label{eq:restrictedcorrelation}
 \langle \prod_{J=1}^M \Phi^{j_J,w_J}_{m_J,\bar m_J} (z_J) \rangle^{\Sigma_0}_\Omega = 
\int \prod_{J=1}^M \frac{d \mu_J^2}{|\mu_J|^2} \mu_J^m \bar \mu_J^{\bar m} 
|\hat u|^2 \delta^{(2)}(\hat u-\Omega^{-1}) 
\langle \prod_{J=1}^M \Phi^{j_J,w_J}(\mu_J|z_J) \rangle^{\Sigma_0}
\end{equation} 
where, for a surface $\Sigma= \Sigma_0$ of genus $g=0$, the variable $\hat u$ is given by 
$\hat u = \sum_J \mu_J z_J$. Here $\Omega$ is some complex parameter that we chose when
we wrote the argument of the $\delta$-function. Since the restricted correlation functions 
\eqref{eq:restrictedcorrelation} depend on $\Omega$, we have displayed this parameter as a 
subscript on the right hand side. But the dependence is rather mild. In fact, as one 
can easily see, upon rescaling of $\Omega$ we have 
\begin{equation} \label{eq:rescaling}
 \langle \prod_{J=1}^M \Phi^{j_J,w_J}_{m_J,\bar m_J} (z_J) \rangle^\Sigma_{\eta \Omega}  = 
 \eta ^{-\sum_J (m_J + k w_J/2)} \bar \eta^{-\sum_J
(\bar m_J + k w_J/2)} 
 \langle \prod_{J=1}^M \Phi^{j_J,w_J}_{m_J,\bar m_J} (z_J) \rangle^\Sigma_\Omega \ . 
\end{equation} 
Manipulating (restricted) correlation functions of vertex operators with non-vanishing 
winding numbers $w$ is assisted by the use of the parafermionic representation. In 
particular it allows to derive some very useful identities that we shall review before 
we conclude this section. 

\paragraph{Parafermionic representation of vertex operators.} 
Following e.g.\ \cite{Argurio:2000tb,Hikida:2020kil} we shall represent the \SL2R WZNW 
model in terms of non-compact parafermions and a free boson $\chi$. One may think of the
parafermionic representation as arising from the decomposition $\textit{SL}(2) \sim 
\textit{SL}(2)/\textit{U}(1) \times \textit{U}(1)$ of the target space. The first factor 
$SL(2)/U(1)$ is a geometric realization of parafermions. We denote the 
associated parafermionic primary fields by
\begin{equation}
\Psi^{j}_{m , \bar m}(z,\bar z) \, .
\end{equation}
%see e.g. \cite{Argurio:2000tb,Hikida:2020kil} for more details.
The free boson that comes with
the factor $U(1)$ is denoted by $\chi(z,\bar z)= \chi(z) + \bar \chi(\bar z)$.
This field is normalized such that
\begin{equation}
\chi (z) \chi (w) \sim - \ln (z-w)\
\end{equation}
and it gives rise to the $\mathfrak{u}(1)$ current
\begin{align}
H(z) = i \sqrt{\frac{k}{2}} \partial \chi (z) \,  .
\end{align}
With the parafermionic vertex operators and the free boson, the vertex
operators \eqref{eq:vertexw} of the spectrally flowed states can be written as
\begin{align}
\Psi^{j}_{m , \bar m}(z,\bar z)
= \Phi^{j,w}_{m, \bar m}(z,\bar z)
e^{i\sqrt{\frac{2}{k}} ((m + \frac{kw}{2}) \chi(z) + (\bar m + \frac{kw}{2})
\bar \chi(\bar z) ) } \, . \label{cosetlang}
\end{align}
What makes this parafermionic representation of the \SL2R WZNW
model so useful is the fact that the parafermionic primary fields $\Psi^j_{m , \bar m}$ do not
depend on the spectral flow number $w$. Instead, all the dependence on $w$ is
in the $U(1)$ vertex operator, i.e. in a free field theory where it is rather
easy to deal with.

\paragraph{Concentrating spectral flow.} 
The parafermionic representation of the WZNW model may be employed to show that 
within correlation functions all spectral flow can be concentrated in a single 
insertion point $\xi$ on the worldsheet, 
\begin{equation}\label{eq:paraaux}
\langle \prod_{J=1}^M \Phi^{j_J,w_J}_{m_J,\bar m_J} (z_J) \rangle_\Omega
= \Theta^{(M)}_k
\left(\genfrac{}{}{0pt}{}{j_J,w_J}{{m_J,\bar m_J}} \middle| z_J \right)
\langle v^{(\twind)}(\xi) \, \prod_{J = 1}^M \Phi^{j_J,0}_{m_J,\bar m_J}
(z_J)\rangle_\Omega
\end{equation}
where the winding number $w$ of the vertex operator \eqref{eq:vw} we place at the 
new insertion point $\xi$ is given by
\begin{equation} 
\label{eq:wsumw}
\twind = \sum_{J=1}^M w_J \ .
\end{equation}
The relation we displayed here is certainly also valid for the standard correlation 
function \eqref{eq:correlationmbasis}. The version we stated here, however, used 
the restricted expectation value we defined in eq.\ \eqref{eq:restrictedcorrelation}. 
The prefactor, finally, is 
given by
\begin{equation}
\Theta^{(M)}_k \left(\genfrac{}{}{0pt}{}{j_J,w_J}{{m_J,\bar m_J}}  \middle| z_J \right)
= \left| \frac{\langle  e^{i\sqrt{\frac{k}{2}} w \chi(\xi) } \
\prod_{J =1}^M e^{i\sqrt{\frac{2}{k}} m_J \chi(z_J)} \rangle }
{\langle \prod_{J =1}^M e^{i\sqrt{\frac{2}{k}}(m_J + \frac{k}{2} w_J)
\chi(z_J)} \rangle}\right|^2\ .
\end{equation}
Strictly speaking, the expectation values should be computed without integration 
over the zero mode $\chi_0$ to avoid $U(1)$ charge conservation. But since the total 
charge within the correlators in the numerator and the denominator is the same, it is 
sufficient to take the ratios of the coefficients in front of the $\delta$ functions. 
These factors drop out form the quotient anyway. 
As a consequence, the function $\Theta$ 
is the same for standard and restricted correlators, and it does not depend on the 
parameter $\Omega$.

\section{Embedding Orbifold Correlators into the WZNW Model}
\label{section.4}

In this section we derive a remarkable formula that expresses the orbifold correlation
functions we introduced in section \ref{section.2} in terms of correlators of the WZNW
theory we reviewed in the previous section. We will write down and comment on the
relevant embedding formula in the first subsection before providing a proof in the
second.

\subsection{The embedding formula}

Before we can state the result, we need to
combine the Kac-Wakimoto free field theory with the CFT $\mathcal{X}$. The product
CFT will be denotes by $\mathcal{W}$. Its state space is a product of the state space
for the theory $\mathcal{X}$ with the state spaces of the $\beta\gamma$ ghost system
and linear dilaton $\phi$,
\begin{equation}
    \mathscr{H}_\mathcal{W} = \mathscr{H}_{\mathcal{X}} \otimes \mathscr{H}_{\beta\gamma}
    \otimes \mathscr{H}_{\phi}\ .
\end{equation}
By construction, this state space carries an action of the Virasoro algebra with central
charge $c_\mathcal{W} = 26$. Let us introduce the following vertex operators
\begin{equation}
\Phi^{j,w}_{m,\bar m}(\psi;z) =
\, \psi(z) \,  \Phi^{j,w}_{m,\bar m}(z) \
\end{equation}
for all primary fields $\psi$ of the CFT $\mathcal{X}$. These states satisfy the physical
state condition of bosonic string theory provided that
\begin{equation} \label{eq:psc}
m = \frac{h_\psi - 1}{w} - \frac{j(j-1)}{(k-2)w} - \frac{k}{4} w
\end{equation}
and similarly for $\bar m$, but with $h_\psi$ replaced by $\bar h_\psi$. Since the
physical state condition determines $m$ and $\bar m$ in terms of the other parameters
$j,w$ and $h_\psi$ we shall often omit the subscripts $m,\bar m$ when dealing with fields
that satisfy the physical state condition.

Our main goal in this section is to merely rewrite the orbifold correlation functions
we studied in section \ref{section.2} in terms of correlation functions of vertex operators in the
WZNW model. More precisely we shall prove that for any of the discrete choices of the covering 
map $\Gamma$ the following remarkable formula holds
\begin{align} \label{corrSL2R}
 \langle \prod_{\nu=1}^n V^{(w_\nu)}_{\alpha_\nu} (\psi_\nu;x_\nu)
 \rangle^{S^2}_\lambda & := \sum_\Gamma \sum_{s=1}^\infty \frac{\lambda^s}{s!}
 \int \prod_{a=1}^{s}  d ^2 u_a \times \\[2mm] 
 & \hspace*{2cm} \times \langle \prod_{\nu=1}^n
 \Phi^{j_\nu, w_\nu} (\psi_\nu;z_\nu)  \prod_{a=1}^{s}
 \Phi^{\frac{1}{b^2},2}_{-k,-k} (u_a)  \!\!
 \prod_{j=1}^{R_0+s/2}\!\! \Phi^{\frac{1}{2b^2},-1}_{\frac{k}{2},\frac{k}{2}}(v_j) 
 \rangle^{\Sigma_0}_{q} \nonumber
\end{align}
with the following relation between the parameters $\alpha_\nu$ and $j_\nu$,
\begin{align} \label{js2}
 \alpha_\nu = b \left(j_\nu - \frac{1}{2b^2}  \right)\ .
\end{align}
Note that we have omitted the labels $m,\bar m$ on the vertex operators that are
inserted at $z = z_\nu$ since the physical state condition fixes theses to coincide
with the conformal weights of the fields in the symmetric product orbifold, i.e.\
\begin{equation}  \label{eq:mfromh}
m_\nu  + \frac{kw _\nu}{2} = h_{\alpha_\nu}^{(w_\nu)}(\psi_\nu) \quad , \quad
\bar m_\nu  + \frac{kw _\nu}{2} = \bar h_{\alpha_\nu}^{(w_\nu)}(\psi_\nu)
\end{equation}
as one can check by combining eqs.\ \eqref{eq:psc}, \eqref{js2} with
\eqref{eq:hwalphapsi}. In comparison to the earlier formula \eqref{correv} that
expressed the orbifold  correlators in terms of correlations function of the parent
theory $\mathcal{M}$, the correlators on the right hand side of formula we stated here
are evaluated in the free Kac-Wakimoto field theory we reviewed in the previous section.
The subscript $q$ instructs us to use the restricted correlation function we defined in 
eq.\ \eqref{eq:restrictedcorrelation} with a parameter $\Omega = q = q^\Gamma$ that is 
given by the constant factor $q$ that multiplies the leading term of the numerator 
potential in the derivative $\partial \Gamma$ of the covering map, see eq. 
\eqref{partialGamma}. 

We note that charge conservation for the linear dilaton $\phi$ with background charge
$Q_\phi$ requires that the insertions in the correlator on the right hand side of
eq.\ \eqref{corrSL2R} must obey
\begin{equation} \label{eq:jconstraint}
\sum_{\nu=1}^n j_\nu - \left(\sum_{\nu=1}^n w_\nu - 2R \right) \frac{1}{2b^2} = 1 \ .
\end{equation}
Using eq.\ \eqref{js2} one can easily verify that this constraint on the parameters
$j_\nu$ coincides with the constraint \eqref{eq:alphaconstraint} on the parameters
$\alpha_\nu$.

There are two aspects of formula \eqref{corrSL2R} that deserve to be stressed. On
the one hand, this formula now contains vertex operators insertion at the $R = R_0 +
s/2$ poles $v_j$ of the covering map, in contrast to eq. \eqref{correv}. On the
other hand, the complicated prefactors that multiplied the correlation functions
of the parent theory $\mathcal{M}$ in eq.\ \eqref{correv} do not show up any longer
in eq. \eqref{corrSL2R}. We will understand both aspects through a detailed
calculation below. Later they will then turn out to be decisive for our ability
to rewrite the orbifold correlators in terms of a dual string theory.

{Before we move on to proving eq.\ \eqref{corrSL2R} let us make one 
more comment concerning insertions of the dual screening charge that is associated 
with the marginal operator \eqref{dmarginalop}. As we have explained before, such 
insertions can be used to compute the residues of a second series of poles in the 
correlation functions of the deformed symmetric product orbifold CFT. The proof of
our embedding formula that we are about to discuss easily extends to cases in which 
the dual screening charge is inserted. For $r$ such insertions, the number of 
insertions of the new vertex operators with $w_j=-1$ increases to $R=R_0+(s+r)/2$ 
and of course there are also $r$ additional insertions of the operators 
$\Phi^{1-\frac{1}{b^2},2}_{-k,-k}$ in the correlator on the right hand side  of 
eq.\ \eqref{corrSL2R}. The latter modify the charge condition 
\eqref{eq:jconstraint} by adding a term of the form $r (1 - 3/2b^2)$ on the left hand 
side of the equation. We note that thereby we are able to explore all the poles 
of the correlation functions in the symmetric product orbifold CFT through the 
embedding into the Kac-Wakimoto representation.}

\subsection{Proof of the embedding formula}
\label{section.4.2}

In order to prove this results we will start with the correlator in the integrand
on the right hand side of eq.\ \eqref{corrSL2R}. Let us define
\begin{equation}
C^{(n;s)}_{g=0}\left(\genfrac{}{}{0pt}{}{j_\nu,w_\nu}{{m_\nu,\bar m_\nu}}  \middle| \,
{\genfrac{}{}{0pt}{}{z_\nu,u_a}{v_j,q}} \right) :=
\langle \prod_{\nu=1}^n
 \Phi^{j_\nu, w_\nu}_{m_\nu,\bar m_\nu} (z_\nu)  \prod_{a=1}^{s}
 \Phi^{\frac{1}{b^2},2}_{-k,-k} (u_a)  \!\! \prod_{j=1}^{R_0+s/2}\!\!
 \Phi^{\frac{1}{2b^2},-1}_{\frac{k}{2},\frac{k}{2}}(v_j) \rangle^{\Sigma_0}_{q} \, .
\end{equation} 
Our goal now is to show that this correlator agrees with the integrand on the
right hand side of eq.\ \eqref{deformedcorr}. The correlator $C$ is computed in
the free field theory \eqref{eq:S0} that contains a $\beta\gamma$ system in
addition to the linear dilaton $\phi$. The correlator in the integrand of
eq.\ \eqref{deformedcorr}, on the other hand, is evaluated in the linear
dilaton theory of the field $\varphi$. To relate the two expressions,
including all the prefactors in \eqref{deformedcorr}, we need to integrate
out the $\beta\gamma$ system and pass from the linear dilaton $\phi$ of the
WZNW model to the linear dilaton $\varphi$ of the orbifold. In doing so, we
shall eventually follow the strategy that we used in \cite{Hikida:2008pe}
to prove the FZZ duality conjecture. But before we can do so, we need to
massage the starting point of the calculation.

\paragraph{Step 1: Collecting the spectral flow.} In a first step we apply the parafermionic
representation, see subsection \ref{section.3.2}, in order to collect all the spectral
flow in a single insertion point which we denote by $\xi$. The combined spectral flow
from all the individual vertex operators adds up to
\begin{equation}
\twind = \sum_{\nu=1}^n w_\nu + 2s - R = n + R + s - 2 \ .
\end{equation}
Hence, by inserting our general formula \eqref{eq:paraaux}, we can write our
correlator in the from
\begin{equation}
C^{(n;s)}_{g=0}\left({\genfrac{}{}{0pt}{}{j_\nu,w_\nu} {m_\nu,\bar m_\nu}} \middle| \,
{\genfrac{}{}{0pt}{}{z_\nu,u_a}{v_j,q}} \right) = \left|\Theta^{(n,R)}_k
\left(\genfrac{}{}{0pt}{}{j_\nu,w_\nu}{m_\nu,\bar m_\nu} \middle| \genfrac{}{}{0pt}{}{z_\nu,u_a}{v_j,\xi} \right)\right|^2
\mathcal{C}^{(n;s)}_{g=0}\left(\genfrac{}{}{0pt}{}{j_\nu,w_\nu}{{m_\nu,\bar m_\nu}}  \middle| \,
{\genfrac{}{}{0pt}{}{z_\nu, u_a}{v_j,q,\xi}} \right)
\end{equation}
where $\mathcal{C}$ denotes the correlation function
\begin{equation}  \label{eq:Cwxi}
\mathcal{C}^{(n;s)}_{g=0}\left(\genfrac{}{}{0pt}{}{j_\nu,w_\nu}{{m_\nu,\bar m_\nu}}  \middle| \,
{\genfrac{}{}{0pt}{}{z_\nu,u_a}{v_j,q,\xi}} \right) :=
\langle v^{(\twind)}(\xi) \, \prod_{\nu=1}^n
 \Phi^{j_\nu, 0}_{m_\nu,\bar m_\nu} (z_\nu)  \prod_{a=1}^{s}
 \Phi^{\frac{1}{b^2},0}_{-k,-k} (u_a)  \!\! \prod_{j=1}^{R_0+s/2}\!\!
 \Phi^{\frac{1}{2b^2},0}_{\frac{k}{2},\frac{k}{2}}(v_j) \rangle^{\Sigma_0}_{q}
\end{equation}
and the factor $\Theta$ is given by the following ratio of correlation functions of the
free field $\chi$
\begin{align} \label{eq:ThetanR}
\Theta^{(n,R)}_k \left(\genfrac{}{}{0pt}{}{j_\nu,w_\nu}{m_\nu,\bar m_\nu} \middle|
{\genfrac{}{}{0pt}{}{z_\nu,u_a}{v_i,\xi}}\right)
& =  \frac{\langle  e^{i\sqrt{\frac{k}{2}} \twind \chi(\xi) } \
\prod_{\nu =1}^n e^{i\sqrt{\frac{2}{k}} m_\nu \chi(z_\nu)} \prod_{a=1}^s e^{-i\sqrt{2k} \chi(u_a)}
\prod_{i=1}^R e^{i\sqrt{\frac{k}{2}}\chi(v_i)}\rangle }{\langle \prod_{\nu =1}^n e^{i\sqrt{\frac{2}{k}}(m_\nu + \frac{k}{2} w_\nu)
\chi(z_\nu)} \rangle} \, \\[4mm]
& =  \frac{\prod^{n+R+s}_{I < J} (z_{IJ})^{\frac{2}{k}m_I m_J}
\prod^{n+R+s}_J (z_J - \xi)^{ \twind m_J}}{\prod_{\mu < \nu }^n
(z_{\mu\nu})^{\frac{2}{k}(m_\mu + \frac{k}{2} w_\mu) (m_\nu + \frac{k}{2} w_\nu)}}
\ .    \label{eq:Theta2}
\end{align}
Here we have combined the labels $\mu = 1, \dots, n$, $a = 1, \dots, s$ and $i=1, \dots, R$
into a single label $J$ that runs through $J=1, \dots, n+R+s$ and we set $z_{n+a} = u_a$ and
$z_{n+s+j} = v_j $. At the same time we also set
\begin{equation}
\label{eq:wJmJ}
w_J = \left\{
\begin{array}{lll} w_\nu &\mbox{for} & \nu = J \leq n \\[2mm]
2  &\mbox{for} & n < J \leq n+s \\[2mm]
-1 \  & \mbox{for} & n+s < J \end{array} \right.
\quad , \quad
m_J = \left\{
\begin{array}{lll} m_\nu &\mbox{for} & \nu = J \leq n \\[2mm]
-k  &\mbox{for} & n < J \leq n+s \\[2mm]
k/2 \  & \mbox{for} & n+s < J \end{array} \right.
\end{equation}
and similarly for $\bar m_J$. Note that the denominator in the second line contains trivial
insertions at the points $z = z_J, J > n,$ on the worldsheet since $m_J + w_J k/2 = 0$ for
all $J>n$. For later use we also introduce
\begin{equation}
\label{eq:jJaJ}
j_J = \left\{
\begin{array}{lll} j_\nu &\mbox{for} & \nu = J \leq n \\[2mm]
1/b^2  &\mbox{for} & n < J \leq n+s \\[2mm]
1/2b^2 \  & \mbox{for} & n+s < J \end{array} \right.
\quad , \quad
\alpha_J = \left\{
\begin{array}{lll} \alpha_\nu &\mbox{for} & \nu = J \leq n \\[2mm]
1/2b \  &\mbox{for} & n < J \leq n+s \\[2mm]
0  & \mbox{for} & n+s < J \end{array} \right. \ .
\end{equation}
After moving all the spectral flow into one single insertion points at $z=\xi$ through
the parafermionic representation our task now is to compute the correlation function
\eqref{eq:Cwxi}.

\paragraph{Step 2: Inverting the spectral flow.} The correlator \eqref{eq:Cwxi} is still
not quite in the form of the correlation function that we dealt with in \cite{Hikida:2008pe}.
In the case at hand, the insertion of the operators $v^{(w)}(\xi)$ forces $\beta (z)$ to possess
a pole of order $w$ at the point $z = \xi$ on the worldsheet. The analysis in \cite{Hikida:2007tq,
Hikida:2008pe}, on the other hand, deals with the case where $\beta (z)$ has a zero at $z = \xi$.
We can relate the two different signs of the spectral flow through the following non-linear
transformation of the $\beta\gamma$ system,
\begin{align}  \label{eq:hatbetagamma}
\hat \gamma(z) = \gamma^{-1}(z) \quad , \quad \hat \beta(z) =
- \gamma(z)^{2} \beta(z) + 2 \partial \gamma(z) \ .
\end{align}
One can verify directly that the new fields $\hat\gamma$ and $\hat \beta$ satisfy the same
operator products as the original fields $\gamma$ and $\beta$. Alternatively, we can also
arrive at the new fields $\hat \gamma$ and $\hat \beta$ through the bosonization of the
$\beta\gamma$ system. Let us recall that the original $\beta \gamma$ system can be
written as
\begin{align}
 \beta (z) \simeq e^{- \bx(z) + \by (z)} \partial \by (z) \, , \quad \gamma (z) \simeq
 e^{\bx(z) - \by(z)}
\end{align}
in terms of two decoupled linear dilaton fields $\bx,\by$ with operator products
\begin{align}
 \bx (z) \bx(0) \sim - \ln z \, , \quad \by (z) \by (0) \sim \ln z \,
\end{align}
and background charges $Q_\bx = -\frac{1}{2}$ and $Q_\by = \frac{1}{2}$, respectively. In
terms of the bosonic fields $\chi^0,\chi^1$ the spectrally flowed vacua of the $\beta \gamma$
system can be expressed as
\begin{align}
v_{(\beta, \gamma)}^{(w)} (z) \simeq e^{- w \bx (z)} \, . \label{vbg}
\end{align}
We can now pass to the new fields $\hat \gamma$ and $\hat\beta$ through a simple
reflection of the fields $\bx, \by$, i.e.\ in the bosonized variables the fields $\hat
\beta$ and $\hat \gamma$ are given by
\begin{align} \label{bosonization}
\hat \beta (z) \simeq - e^{\bx (z) - \by (z)} \partial \by(z) \, , \quad
\hat \gamma  (z) \simeq e^{- \bx(z) + \by (z)} \, .
\end{align}
If we apply the same reflection to the bosonization formula \eqref{vbg} for the spectrally
flowed vacua we find that
$$v^{(w)}_{(\beta,\gamma)}= v^{(-w)}_{(\hat\beta,\hat\gamma)}\ . $$
In conclusion we have shown that the correlator \eqref{eq:Cwxi} can be rewritten as
\begin{align}  \label{eq:Cwxihat}
\begin{aligned}
&\mathcal{C}^{(n;s)}_{g=0}\left({\genfrac{}{}{0pt}{}{j_\nu,w_\nu}{m_\nu,\bar m_\nu}} \middle| \, 
{\genfrac{}{}{0pt}{}{z_\nu,u_a}{v_j,q,\xi}} \right) \\ 
&:= q^{- \sum_\nu (m_\nu + \bar m_\nu + k w_\nu) }
\langle \hat v^{(\twind)}(\xi) \, \prod_{\nu=1}^n
 \hat\Phi^{j_\nu, 0}_{m_\nu,\bar m_\nu} (z_\nu)  \prod_{a=1}^{s}
 \hat \Phi^{\frac{1}{b^2},0}_{-k,-k} (u_a)  \!\! \prod_{j=1}^{R_0+s/2}\!\!
 \hat\Phi^{\frac{1}{2b^2},0}_{\frac{k}{2},\frac{k}{2}}(v_j) \rangle^{\Sigma_0}_{1}
 \end{aligned}
\end{align}
where the vertex operators $\hat \Phi$ 
have the same form as in eq.\ \eqref{eq:vertex} but with $\gamma$ replaced by $\hat \gamma^{-1}$. 
The spectral flow carrying operator $\hat v$ 
is
\begin{align} \label{hatu}
 \hat v^{(w)} = v^{(-w)}_{(\hat\beta,\hat \gamma)} e^{-\frac{w}{b}\phi}\ . 
\end{align}

\paragraph{Step 3: Integrating out the $\beta\gamma$ system.} After completing the
previous two steps, we are precisely in the setup we addressed in \cite{Hikida:2008pe}.
The analysis of \cite{Hikida:2007tq,Hikida:2008pe} is based on passing from the $m$-basis to the so-called $\mu$-basis, see our discussion in section \ref{section.3.1} and in particular eq.\ \eqref{eq:mfrommu}. 
In order to undo the sign change of the 
indices $m$ and $\bar{m}$ we can invert $\mu$ and $\bar \mu$ to obtain 
\begin{align}
 \hat \Phi^{j,0}_{m , \bar m} (z)=
 \int \frac{d^2 \mu}{|\mu|^2} \mu^m \bar \mu^{\bar m} \hat \Phi^{j} (\mu|z) \, , \quad
\hat \Phi^j (\mu |z) = |\mu|^{2j} e^{\frac{1}{\mu} \hat\gamma (z) - \frac{1}{\bar \mu}
\hat {\bar \gamma} (\bar z)} e^{2 b j \phi (z, \bar z)} \ . \label{mubasis}
\end{align}
Putting all this together we arrive at the following expression for our correlation function
\begin{align} \label{Cmup}
\begin{aligned}
&\mathcal{C}^{(n;s)}_{g=0}\left({\genfrac{}{}{0pt}{}{j_\nu,w_\nu}{m_\nu,\bar m_\nu}} \middle| \,
{\genfrac{}{}{0pt}{}{z_\nu,u_a}{v_j,q,\xi}} \right)  \\&=q^{- \sum_\nu (m_\nu + \bar m_\nu + k w_\nu) }
\int \prod_{I=1}^{n+R+s} \frac{d^2 \mu_I}{|\mu_I|^2} \mu_I^{m_I} \bar \mu_I^{\bar m_I}\,
 \delta^{(2)}(\hat u - 1) \ 
\langle \hat v^{(w)} (\xi)
\prod_{I=1}^{n + R+s}  \hat \Phi^{j_I} (\mu_I | z_I) 
 \rangle
 \end{aligned}
\end{align}
with the parameters $j_{I}$ as introduced in eq.\ \eqref{eq:jJaJ}. The $\delta$-function 
in the integrand implements the construction of the restricted correlation function, see 
eq.\ \eqref{eq:restrictedcorrelation}. The variable $\hat u$ is given by $\hat u = \sum_I 
z_I/\mu_I$. Following \cite{Hikida:2007tq,Hikida:2008pe} we perform the path integral over 
$\hat \gamma$ to obtain the a constraint on $\bar \partial \hat \beta$. After 
integration with respect to the worldsheet coordinate this constraint reads
\begin{align}
\hat \beta (z) = \sum_{I =1}^{n + R+s} \frac{1/\mu_I}{z - z_I} =
\hat u \frac{(z - \xi)^{n + R + s -2}}
{\prod_{I=1}^{n + R +s}(z - z_I)} \, . \label{mu2up}
\end{align}
Note that the residues are given by $1/\mu_I$ instead of $\mu_I$ because of the way
we have introduced our $\mu$-basis. The second equality arises from the insertion of
the operator $\hat v^{(w)}(\xi)$ which forces $\hat \beta$ to have a zero of order
$w$ at $z=\xi$. But this equality can only hold provided that the parameters $\mu_I$
satisfy the following $n + R + s -1$  constraint equations
\begin{align}  \label{eq:muconstraints}
\sum_{I=1}^{n+R+s} \frac{1/\mu_I}{(\xi - z_I)^p} = 0 \, , \quad
p = 0,\dots,w\, .
\end{align}
Since we have integrated  $\hat \beta (z) , \hat \gamma (z)$ the  correlation function
is now obtained as a path-integral only over the field $\phi(z, \bar z)$. As in
\cite{Hikida:2007tq}, we shift the field $\phi(z,\bar z)$ using the prescription
\begin{align}\label{eq:phishift}
 \varphi (z , \bar z) = \phi (z , \bar z)
 - \frac{1}{2 b} \left( w \ln |z - \xi|^2  - \sum_{I =1}^{n + R +s }
\ln |z - z_I| ^2  - \ln |\rho(z)|^2 \right) \, .
\end{align}
The correlation function \eqref{Cmup} is now
given by, see \cite{Hikida:2007tq} for details,
\begin{align}
\begin{aligned}
&\mathcal{C}^{(n;s)}_{g=0}\left({\genfrac{}{}{0pt}{}{j_\nu,w_\nu}{m_\nu,\bar m_\nu}} \middle| \,
{\genfrac{}{}{0pt}{}{z_\nu,u_a}{v_j,q,\xi}} \right)  = q^{- \sum_\nu (m_\nu + \bar m_\nu + k w_\nu) } \int \prod_{I=1}^{n+R+s}
\frac{d^2 \mu_I}{|\mu_I|^2} \mu_I^{m_I} \bar \mu_I^{\bar m_I}\, \\[2mm]
& \quad  \times  
\prod_{p=0}^{n+R+s-2} \delta^2 \left( \sum_{\nu=1}^{n+R+s} \frac{1/\mu_I}{(\xi - z_I)^p}\right)\delta^{(2)}(\hat u - 1)\,   \prod_{I < J} |z_{I J}|^{\frac{1}{b^2}}  \
\langle \prod_{I=1}^{n+R+s}
V_{\alpha_I}(z_I)\ \rangle^{\Sigma_0} \,
\end{aligned}
\end{align}
where $V_\alpha$ denotes the vertex operators in the linear dilaton theory that
were defined in eq.\ \eqref{eq:Valpha} and the correlation function on the right
hand side is to be evaluated in the linear dilaton theory of the field $\varphi$
that we introduced in our discussion of the parent CFT, see section \ref{section.2}. 
In particular, the shift that we defined in eq.\ \eqref{eq:phishift} implies that the 
new field $\varphi$ has background charge $Q = Q_\varphi$, just as in eq.\ \eqref{eq:Qvarphi}.

In order to explicitly pass back form the $\mu$-
to the $m$-basis, we want to perform the integrals over all the parameters $\mu_I$.
This is helped by the fact that the $n+R+s$ integrations are constrained by $n+R+s$
conditions in the arguments of the $\delta$-functions. Using the following result for
the Jacobian, see \cite{Ribault:2005wp,Ribault:2005ms,Hikida:2008pe})
\begin{align}
\prod_{I=1}^{n+R+s} \frac{d^2 \mu_I}{|\mu_I|^2}  \prod_{p=0}^{n + R + s -2 }
\delta^2 \left(\sum_{J=1}^{n + R+s} \frac{1/\mu_J}{(\xi - z_J)^p} \right)  =
\frac{d^2 \hat u}{| \hat u |^{2 w + 4}} \, \prod_{I < J}|z_{I J}|^2 \, ,
\end{align}
we find
\begin{align}
\label{eq:Cnfinal}
\mathcal{C}^{(n;s)}_{g=0}\left({\genfrac{}{}{0pt}{}{j_\nu,w_\nu}{m_\nu,\bar m_\nu}} \middle| \,
{\genfrac{}{}{0pt}{}{z_\nu,u_a}{v_j,q,\xi}} \right) =  |\theta|^2
\left[  \prod_{I=1}^{n+ R+s}\mu_I^{m_I} \bar \mu_I^{\bar m_I}  \right]
\, \langle \prod_{\nu=1}^{n} V_{\alpha_\nu}(z_\nu)
\prod_{a=1}^s V_{\frac{1}{2b}}(u_a) \rangle^{\Sigma_0}
\end{align}
with a factor $\theta$ that is given by
\begin{align} \label{eq:theta}
|\theta |^2 =  q^{- \sum_\nu (m_\nu + \bar m_\nu + k w_\nu) }  \prod_{I < J} |z_{I J}|^{k} \, .
\end{align}
Here we have converted the product over vertex operators in the linear dilaton theory
into three separate products over $\nu = 1, \dots, n$, $a=1, \dots, s$ and $j = 1, \dots, R$.
Since the parameters $\alpha_J = 0$ for $n+s < J$, see eq.\ \eqref{eq:jJaJ}, the vertex
operator insertions at the poles $z = z_j$ of the covering map $\Gamma$ are trivial. We
note that the resulting correlator coincides with the correlator on the right hand side
of eq. \eqref{deformedcorr}.

In order to compare the coefficients, we exploit the parameters $\mu_I$ can be read off
from eq.\ \eqref{mu2up} to take the form
\begin{align} \label{eq:muI}
\mu_I =  \frac{\prod_{J\neq I}^{n + R+s} z_{I J} }{(z_I - \xi)^{\twind} }\ ,
\end{align}
where we applied the regularization scheme that sets $\lim_{z \to w} \ln |z - w|^2 =
- \ln |\rho (w)|^2 = 0$. After multiplication with the factor $\Theta$ that was
defined in eq.\ \eqref{eq:ThetanR} we finally arrive at  
\begin{equation}
C^{(n;s)}_{g=0}\left({\genfrac{}{}{0pt}{}{j_\nu,w_\nu} {m_\nu,\bar m_\nu}} \middle| \,
{\genfrac{}{}{0pt}{}{z_\nu,u_a}{v_j,q}} \right) = \,  \vartheta^{(n,R)}_{g,k}
\left(\genfrac{}{}{0pt}{}{j_\nu,w_\nu}{m_\nu,\bar m_\nu}\middle|  \,
{\genfrac{}{}{0pt}{}{z_\nu,u_a}{v_j,q}} \right) \, \langle
\prod_{\nu=1}^{n} V_{\alpha_\nu}(z_\nu)  \prod_{a=1}^s V_{\frac{1}{2b}}(u_a) \rangle^{\Sigma_0}
\end{equation}
with a prefactor $\vartheta$ that is given by
\begin{equation} \label{eq:varthetaresult}
\vartheta^{(n,R)}_{g,k} \left(\genfrac{}{}{0pt}{}{j_\nu,w_\nu}{m_\nu,\bar m_\nu}
\middle|  \,
{\genfrac{}{}{0pt}{}{z_\nu,u_a}{v_j,q}} \right) = |\theta|^2
\left|\Theta^{(n,R)}_k \left(\genfrac{}{}{0pt}{}{j_\nu,w_\nu}{m_\nu,\bar m_\nu} \middle|  \,
{\genfrac{}{}{0pt}{}{z_\nu,u_a}{v_j,q,\xi}}\right)\right|^2\
\left[  \prod_{I=1}^{n+ R+s}\mu_I^{m_I} \bar \mu_I^{\bar m_I}  \right] \ .
\end{equation}
Using the explicit expressions in eqs.\ \eqref{eq:theta}, \eqref{eq:ThetanR} and
\eqref{eq:muI} one can check straightforwardly that the function $\vartheta$ coincides
with the prefactor in the integrand of eq.\ \eqref{deformedcorr}, i.e. 
\begin{align}
\vartheta^{(n,R)}_{g,k} \left(\genfrac{}{}{0pt}{}{j_\nu,w_\nu}{m_\nu,\bar m_\nu}
\middle|  \, {\genfrac{}{}{0pt}{}{z_\nu,u_a}{v_j,q}} \right)
 = \prod_{\nu=1}^n
(\tilde a_\nu^\Gamma)^{-h_{\alpha_\nu}^{(w_\nu)} + \frac{k}{4} (w_\nu -1)} (\bar{\tilde a}_\nu^\Gamma)^{-\bar h_{\alpha_\nu}^{(w_\nu)} + \frac{k}{4} (w_\nu -1)}
\prod_{a=1}^s |\tilde b^\Gamma_a|^{\frac{k}{2}}\prod_{j=1}^{R_0+s/2}
|\xi_j^\Gamma|^{-k} \ .
\end{align}
This is a fairly remarkable result that finally establishes the formula \eqref{corrSL2R} 
for the correlators of the perturbed symmetric product orbifold in terms of a correlator 
in the WZNW model. Let us stress however, that our formula is purely within CFT. In fact, 
on both sides of the equation the fields are inserted at specific points on the worldsheet. 
This is to be contrasted with expressions for string amplitudes in which insertion points 
of worldsheet fields are integrated over. It is the task of the next section to uncover a 
relation of the CFT correlators with string scattering amplitudes. Our formula \eqref{corrSL2R}
holds the key for such a string theoretic formulation.

\section{The Uplift to String Theory}
\label{section.5}

After rewriting the correlators of the symmetric product orbifold in terms of correlation
functions of operators in some $c_\mathcal{W}=26$ CFT we are now well prepared uplift the
latter to scattering amplitudes in a bosonic string theory. In the first subsection we
take a closer look at certain vertex operators of $\mathcal{W}$. Most importantly, we shall 
discuss two integrated vertex operators that commute with the currents and the stress 
tensor of the Kac-Wakimoto free field theory. These can be added as interaction terms 
to the free field theory while preserving both current algebra and Virasoro symmetry. 
The action of this dual string theory will be spelled out in the second subsection. 
For this theory we shall then analyse tree level scattering amplitudes and show that 
they agree with the correlation functions of the symmetric product orbifold in the 
planar limit. 

\subsection{Screening charges and vertex operators} 

\paragraph{Screening charge in the $w=-1$ sector.}
A screening charge is a (scale invariant) integrated vertex operator that
commutes with all three currents. The conventional screening charges of the 
\SL2R WZNW model are constructed 
from fields in the vacuum sector with winding number $w=0$ of the current algebra. 
But these won't play any role below. Instead we shall now discuss screening charges 
that carry non-vanishing winding number, starting with $w=-1$. As was observed in 
\cite{Giveon:2019gfk} already the following object preserves the current algebra 
symmetry 
\begin{equation} \label{eq:intm}
S^- = - \int d^2v \, \Phi^{\frac{1}{2b^2},-1}_{\frac{k}{2},\frac{k}{2}}(v,\bar v)
\sim - \int d^2v \ [\gamma^{1-k}_1\bar \gamma^{1-k}_1
|w=-1\rangle_{(\beta,\gamma)}](v,\bar v)\  e^{\frac{2}{b}\phi (v,\bar v)}\ .
\end{equation}
On the right hand side we have written out the construction of the relevant
field $\Phi$, see eq.\ \eqref{eq:vertexw}. Recall that $\gamma_1$ is a Fourier
mode of the field $\gamma(z)$. In order to verify that the integrated operator
is indeed invariant under rescalings of the worldsheet, we note that the flowed
vacuum $|w=-1\rangle_{(\beta,\gamma)}$ of the $\beta\gamma$
system with winding number $w=-1$ has weight $h_{w=-1} = -1$.  By application of
the $1-k$ lowering operators $\gamma_1$ and $\bar \gamma_1$ this weights are
lowered to $(h,\bar h) = (k-2,k-2) = (1/b^2,1/b^2)$. When combined with the vertex
operator $\exp (2\phi/b)$ of weight $h = 1 - 1/b^2$ the resulting integrand indeed
has one unit of left and right moving weight. It is also easy to show that the
integral commutes with the currents.

The interaction term we displayed in eq.\ \eqref{eq:intm} is a close relative of a
more conventional screening charge that was first found by Bershadsky and Ooguri
\cite{Bershadsky:1989mf},
\begin{equation}\label{eq:primeint}
\check{Q} = - \int d ^2 v \ \beta^{k-2}(v) \bar
\beta^{k-2}(\bar v) e^{\frac{2}{b} \phi(v,\bar v)} \, .
\end{equation}
Indeed, the dependence on the linear dilaton $\phi$ is the same in both screening
charges. In the ghost sector, the states $\gamma_1^{1-k}|w=-1\rangle$ that describe
the chiral half of the integrand in eq.\ \eqref{eq:intm} are close relatives of the
states $\beta_{-1}^{k-2}|w=0\rangle$ that appear as chiral half of the ghost sector
in the integrand of screening charge \eqref{eq:primeint}.

Before we conclude this short discussion of the screening charge \eqref{eq:intm} we
want to briefly display the operator product of the integrand with the field $\gamma$
which reads
\begin{equation}
\gamma(z) \, \Phi^{\frac{1}{2b^2},-1}_{\frac{k}{2},\frac{k}{2}}(v,\bar v)
\sim \frac{1}{z-v}\,
:\gamma \Phi^{\frac{1}{2b^2},-1}_{\frac{k}{2},\frac{k}{2}}\!\! :(v,\bar v) + \dots
\end{equation}
i.e.\ the field $\gamma$ has a first order pole at the insertion point $(v,\bar v)$.
We have already pointed at some similarities between $\gamma$ and the covering map
at the end of the previous paragraph. Since the covering map does indeed possess
poles, we expect the screening charge $S^-$ to be relevant for our theory. And
indeed one can see that the corresponding local field does appear inside the
correlator on the right hand side of our equation \eqref{corrSL2R}. We will
get back to this point in the next subsection.

\paragraph{Screening charge in the $w=1$ sector.}
Having discussed a screening charge with winding number $w=-1$ it seems natural to 
turn to the opposite direction and look for a screening charge with $w=1$. Such a 
screening charge does indeed exist \cite{Giveon:2019gfk} and is given by 
\begin{equation}
S^+ = - \int d^2 u \Phi^{\frac{1}{2b^2},1}_{-\frac{k}{2},-\frac{k}{2}}(u,\bar u)
= -  \int d^2u \ [\gamma_{-1}\bar \gamma_{-1}
|w=1\rangle_{(\beta,\gamma)}](u,\bar u) \ .
\end{equation}
Note that this operator does not depend on the bosonic field $\phi$. Since the
spectrally flowed vacuum $|w=1\rangle$ has vanishing conformal weight, the
integral is obviously invariant under scale transformations.

In addition, it also commutes with the currents. To keep our discussion self-contained
let us briefly prove this statement here. The most important observation concerns the 
operator product expansion of the field $J^+(w) = - \beta(w)$ with the operator in the 
integrand of the screening charge, or rather its chiral half,
\begin{eqnarray}
J^+(z) \, [\gamma_{-1} |w=1\rangle_{(\beta,\gamma)}](u) & \sim &
-\frac{1}{(z-u)^{2}} [\beta_1\gamma_{-1} |w=1\rangle_{(\beta,\gamma)}](u)
- \frac{1}{z-u} [\beta_0 \gamma_{-1} |w=1\rangle_{(\beta,\gamma)}](u) + \dots
\nonumber \\[2mm]
& \sim & - \frac{1}{(z-u)^{2}} [|w=1\rangle_{(\beta,\gamma)}](u)
+ \frac{1}{z-u} [L_{-1} |w=1\rangle_{(\beta,\gamma)}](u) + \dots \, .
\end{eqnarray}
In the first term we have used that $\beta_n|w=1\rangle = 0$ for $n > 0$ along
with the usual commutation relations in the $\beta\gamma$ system. To show that the
second term involves a derivative we used
$$
L^{(\beta,\gamma)}_{-1} = \sum_{n>0} \left(n \beta_{-n-1} \gamma_n - n
\gamma_{-n} \beta_{n-1}\right)$$
along with $\gamma_n|w=1\rangle = 0 $ for $n \geq 0$. The final result of this
short calculation can be expressed more elegantly as
\begin{equation}
J^+(z) \, \Phi^{\frac{1}{2b^2},1}_{-\frac{k}{2},-\frac{k}{2}}(u,\bar u)
\sim - \frac{1}{(z-u)^2}\ v^{(w=1)}_{(\beta,\gamma)}(u) + \frac{1}{z-u}
\partial_z v^{(w=1)}_{(\beta,\gamma)}(u)
\end{equation}
where $v^{(w=1)}_{(\beta,\gamma)}$ denotes the fields that corresponds to the
ground state $|w=1\rangle_{(\beta,\gamma)}$ of the $\beta\gamma$ system. Note
that the first order pole multiplies a derivative field. This establishes that
the current $J^+$ commutes with our screening charge. For the other two currents
one may proceed along similar lines.%

The screening charge $S^+$ has a number of remarkable properties that distinguishes it
from other screening charges, and in particular from $S^-$. The first property we want
to look at concerns the behavior of the field $\gamma$ in the vicinity of the unintegrated
operator. Since this operator has $w=1$, it produces a simple zero of $\gamma(z)$ at the
insertion point $u$, 
\begin{align}
 \label{gammaQ}
\begin{aligned}
\gamma (z) \Phi^{\frac{1}{2b^2},1}_{- \frac{k}{2},-\frac{k}{2}} (u)  =
 \gamma (z)  [\gamma_{-1} \bar \gamma_{-1}| w=1 \rangle_{(\beta , \gamma)}] (u)
 =  (z-u) [(\gamma_{-1})^2 \bar \gamma_{-1}| w=1 \rangle_{(\beta , \gamma)}] (u) +
 \mathcal{O} ((z-u)^2) \, .
\end{aligned}
\end{align}
But there is a more remarkable property of the screening charge $S^+$ that can be 
understood from our calculations in the previous section: An insertion of $S^-$ into 
certain correlation functions with maximal violation of winding number is actually 
trivial. The precise statement is 
\begin{equation} \label{eq:triviality}
\langle \prod_{\nu=1}^n
 \Phi^{j_\nu, w_\nu} (\psi_\nu;z_\nu)\,
 \Phi^{\frac{1}{2b^2},1}_{-\frac{k}{2},-\frac{k}{2}} (u)
 \, \prod_{j=1}^{R_0} \Phi^{\frac{1}{2b^2},-1}_{\frac{k}{2},\frac{k}{2}}(v_j)
 \rangle^{\Sigma_0}_q = \langle \prod_{\nu=1}^n
 \Phi^{j_\nu, w_\nu} (\psi_\nu;z_\nu)
 \prod_{j=1}^{R_0} \Phi^{\frac{1}{2b^2},-1}_{\frac{k}{2},\frac{k}{2}}(v_j)
 \rangle^{\Sigma_0}_q \ .
\end{equation}
This equation is a consequence of our result \eqref{corrSL2R} along with eq.\ \eqref{correv}.
Indeed, we can obtain eq.\ \eqref{eq:triviality} from the previous analysis if we set $s=0$
and replace $n$ by $n+1$ with an additional vertex operator inserted at $z_{0}=u$ whose
quantum numbers are
\begin{align} \label{parameters}
 j_{0} = \frac{1}{2 b^2} \, , \quad w_0 = 1 \, ,\quad m_0 = \bar m_0 = - \frac{k}{2} \, .
\end{align}
Formula \eqref{corrSL2R} allows us to evaluate this correlator in terms of some $n+1$-point
function in the symmetric product orbifold. The latter contains a twist field that is associated
to a cycle of length $w=1$, which is simply the identity field. Hence the $n+1$-point function 
reduces to an $n$-point function which can be re-expressed as the correlation function on the 
right hand side of equation \eqref{eq:triviality}. In order for all this to work out it
is crucial that the insertion of a WZNW vertex operator with quantum numbers \eqref{parameters}
does neither effect the constraint \eqref{eq:jconstraint} nor the value of the parameter $R_0$.
In addition, one can also verify easily that the two correlation functions we compare in eq.\
\eqref{eq:triviality} give rise to the same prefactors when we evaluate the correlator of
the symmetric product orbifold in the parent theory. Indeed, the values \eqref{parameters}
imply that
\begin{align}
j_0 - \frac{1}{2b^2} =  0 , \quad - m_0 - \frac{k}{4} (w_0 + 1) = - \bar m_0 -
\frac{k}{4} (w_0 + 1) = 0 \, .
\end{align}
When these parameters are inserted into the formulas \eqref{abxi} for the $n+1$-point correlators
one finds
\begin{align}
\begin{aligned}
& \tilde a_\nu^\Gamma = w_\nu a_\nu^\Gamma =   \frac{\prod_{\mu \neq \nu , \mu \geq 1} (z_\nu - z_{\mu})^{w_{\mu} -1}}{ \prod_{j=1}^{R+t/2} (z_\nu - v_j)^2 } \quad (\nu \geq 1) \, ,
\quad \xi_j^\Gamma = \frac{\prod_{\nu  \geq 1 } (v_j - z_\nu)^{w_\nu -1}}{\prod_{i \neq j}(v_j - v_i)^2} \, .
\end{aligned}
\end{align}
In addition, the exponent of $|a_0^\Gamma|$ is trivial. Let us stress that in the expressions
for $a_\nu^\Gamma$ and $\xi_j^\Gamma$ all products run over the indices $\nu$ and $\mu$ start at $\nu,\mu=1$,
i.e.\ they do not include $\nu,\mu =0$. This establishes the triviality property \eqref{eq:triviality}
of our second screening charge $S^-$. Before we close this subsection let us spell out the following 
variant of formula \eqref{eq:triviality} 
\begin{eqnarray} \label{eq:trivialityR}
& & \langle \prod_{\nu=1}^n
 \Phi^{j_\nu, w_\nu} (\psi_\nu;z_\nu)
 \, \Phi^{\frac{1}{2b^2},1}_{-\frac{k}{2},-\frac{k}{2}} (u) \prod_{a=1}^s
 \Phi^{\frac{1}{2b^2},2}_{-k,-k}(u_a)
 \prod_{j=1}^{R_0+s/2} \Phi^{\frac{1}{2b^2},-1}_{\frac{k}{2},\frac{k}{2}}(v_j)
 \rangle^{\Sigma_0} =   \\[2mm] & & \hspace*{4cm}= \langle \prod_{\nu=1}^n
 \Phi^{j_\nu, w_\nu} (\psi_\nu;z_\nu) \prod_{a=1}^s
 \Phi^{\frac{1}{2b^2},2}_{-k,-k}(u_a)
 \prod_{j=1}^{R_0+s/2}\!\! \Phi^{\frac{1}{2b^2},-1}_{\frac{k}{2},\frac{k}{2}}(v_j) \
 \rangle^{\Sigma_0}\ . \nonumber
\end{eqnarray}
This equation will play a very crucial role in the next subsection
where we finally engineer the perturbative string dual.

\paragraph{Deforming the vertex operators.}
In order to pass from unintegrated vertex operators of the CFT to the integrated vertex
operators of the string theory we will need to deform the vertex operators we introduced
in eq.\ \eqref{eq:vertexw}. The deformed operators depend on two `deformation
parameters' $x, \bar x$ as follows
\begin{align}
\Phi^{j , w }_{m, \bar m} (x|z,\bar z) & = N^{j}_{m,\bar m}\,
[ e^{- x \beta_0 - \bar x \bar \beta_0} \gamma^{-j -m}_{-w}
\bar \gamma^{-j - \bar m}_{-w}|w\rangle] (z , \bar z)
e^{2b (j- \frac{w}{2b^2}) \phi (z , \bar z)}   \, .
 \label{xop}
\end{align}
The operator products in the $\beta\gamma$ system state that the field $\beta$ is conjugate
to the field $\gamma$. In particular, the zero mode $\beta_0$ is conjugate to the constant
mode $\gamma_0$ of $\gamma$. Hence, the new $x,\bar x$-dependent factors in the definition
of the deformed vertex operators modify the operator product with the fields $\gamma$, see
\cite{Eberhardt:2019ywk},
\begin{align}
\gamma(z)  \Phi^{j,w}_{m,\bar m}(x|z',\bar z') =
\gamma(z) e^{- x \beta_0 - \bar x \bar \beta_0} \Phi^{j,w}_{m,\bar m} (z',\bar z')
= e^{- x \beta_0 - \bar x \bar\beta_0} e^{x \beta_0} \gamma(z)  e^{- x \beta_0}
\Phi^{j,w}_{m,\bar m} (z',\bar z') \, .
\end{align}
In order to evaluate this further we use
\begin{align}
    e^{x \beta_0} \gamma(z) e^{- x \beta_0} = \gamma (z) + x \, .
\end{align}
Consequently, the operator product between the field $\gamma$ and the deformed vertex
operators reads
\begin{align} \label{eq:gammaPhiw}
 \gamma(z)  \Phi^{j,w}_{m,\bar m} (x|z',\bar z') \sim
 x \, \Phi^{j,w}_{m,\bar m}(x|z',\bar z')  + (z-z')^w
 e^{- x \beta_0-\bar x \bar \beta_0}
 :\gamma \Phi^{j,w}_{m,\bar m}:(x|z',\bar z') + \dots \, .
\end{align}
In order to appreciate this formula, we stress the striking similarity with the behavior
\eqref{Gammadef} of the covering map $\Gamma$. The comparison shows in particular that
the deformation parameter $x$ plays the role of the insertion point in the dual CFT.
This interpretation of the deformation parameter $x$ is also natural since the zero
modes of the $\mathfrak{sl}(2)$ currents generate global conformal transformations of
the CFT on the boundary, with the two modes $J^+_0 = - \beta_0$ and $\bar J^+_0 = - \bar
\beta_0$ corresponding to 2-dimensional infinitesimal translations.
\subsection{The string theory dual}

We are finally in a position to construct the string theory dual to the perturbed symmetric product
orbifold. In order to do so, we go now back to equation \eqref{corrSL2R}. As we pointed out before,
the correlator on the right hand side is a CFT correlator and not yet a string amplitude. It involves
two types of vertex operators that are associated with physical states, namely the operators
$\Phi^{j_\nu,w_\nu}$ and $\Phi^{1/2b^2,-1}$, but these are not integrated over as it is done in
string amplitudes. On the other hand, the operators $\Phi^{1/b^2,2}_{-k,-k}$ are integrated over
but they do not correspond to physical states. Nevertheless we claim that the right hand side of
eq.\ \eqref{corrSL2R}, and hence the correlators of the deformed symmetric product orbifold, are
just a few steps away from an interpretation as a string amplitude in a bosonic string theory that
we are about to construct.

The relevant worldsheet theory is generated by the fields $\beta, \gamma$ and $\phi$ with two
interaction terms that preserve the current algebra symmetries. The action of this theory is
given by
\begin{equation} \label{eq:S}
S[\beta,\gamma,\phi] = S_0[\beta,\gamma,\phi] - \int d^2 v \Phi^{\frac{1}{2b^2},-1}(v,\bar v)
- \mu \int d^2u \Phi^{\frac{1}{2b^2},1}(u, \bar u) \ .
\end{equation}
Here, $S_0$ is the action of the free theory that we spelled out eq.\ \eqref{eq:S0} already.
The two interaction terms involve the two screening charges we introduced in the previous
subsection. Note that we have absorbed a possible coupling constant that one could have
expected in front of the first interaction term by shifting the zero model of the linear 
dilaton $\phi$. In the second interaction term, on the other hand, we kept the coupling 
constant $\mu$. It follows form the discussion in the previous subsection that the action 
$S$ defines an interacting worldsheet CFT with Virasoro central charge $c=26$. The 
interaction terms do not only preserve the Virasoro symmetry but even the full current 
algebra we constructed through eqs.\ \eqref{eq:currents}. By the usual constructions, such 
a worldsheet CFT gives rise to consistent (but non-geometric) bosonic string theory.

Within this bosonic string background we can now construct string amplitudes. To this end, 
we pick a set of $n$ physical states of the worldsheet theory. We will take these to be 
given by
$$ \Phi^{j_\nu,w_\nu}(x_\nu|\psi_\nu;z_\nu,\bar z_\nu) = \psi(z_\nu,\bar z_\nu)
\Phi^{j_\nu,w_\nu}_{m_\nu,\bar m_\nu}(x_\nu|z_\nu,\bar z_\nu)\ $$
where $m_\nu$ and $\bar m_\nu$ are determined by the weights of the state $\psi_\nu$ in the
CFT $\mathcal{X}$ through eq.\ \eqref{eq:mfromh} and the $x$-dependence of $\Phi$ was introduced
in eq.\ \eqref{xop}. Our claim now is that the associated string amplitude which is obtained by
integrating out the worksheet positions of these vertex operators, reproduces the correlators
of the deformed orbifold theory, i.e. 
\begin{align} \label{amplSL2R}
 \langle \prod_{\nu=1}^n V^{(w_\nu)}_{\alpha_\nu} (\psi_\nu;x_\nu)
 \rangle^{S^2}_\lambda & \simeq \langle \prod_{\nu=1}^n \int d^2z \, \Phi^{j_\nu,w_\nu}
 (x_\nu|\psi_\nu; z,\bar z) \rangle_{\mu = \sqrt{\lambda}} \ .
\end{align}
The correlator of integrated vertex operators on the right hand side is computed in the
theory \eqref{eq:S} with coupling parameter $\mu$ related to the deformation parameter
$\lambda$ of the dual CFT by $\lambda = \mu^2$. As usual, making sense of the right hand 
side requires to fix a frame in which three insertion points are fixed to $(0,1,\infty)$
so that the integration only extends over the remaining $n-3$ insertion points. Finally,
we put a $\simeq$ instead of $=$ because we will prove this equality only up to a 
numerical factor of the form
\begin{equation} \label{eq:Lambda}
\Lambda(\eta) = \Lambda^{w_\nu}_{m_\nu}(\eta) := \eta^{-\sum_\nu (m_\nu + \frac{k w_\nu}{2})} 
\overline{\eta}^{-\sum_\nu
(\bar m_\nu + \frac{k w_\nu}{2})} = \eta^{-\sum_\nu h_\nu} \overline{\eta}^{-\sum_\nu \bar h_\nu}  
\end{equation} 
for some complex parameter $\eta$. On the right hand side we have expressed the 
exponents in terms of the conformal weights \eqref{eq:mfromh} of the field in the 
symmetric product orbifold. This shows that any factor for the form $\Lambda(\eta)$ 
can be absorbed into a rescaling of the insertion points $x_\nu$ of the symmetric 
product orbifold. As such, it does now contain any dynamical information about the 
CFT beyond the scaling weight. Consequently, by working only modulo factors of the 
form \eqref{eq:Lambda} we do not loose any non-trivial content of the relation.  
 
The proof of formula \eqref{amplSL2R} relies heavily on the localization properties of
the path integral over $\gamma$ in the worldsheet correlation functions. These assert 
that in correlators with maximal violation of winding number, the worldsheet field 
$\gamma$ localizes to the branching function $\Gamma$ that we introduced in eq.
\eqref{Gammadef} while discussing the perturbative expansion of the symmetric product 
orbifold,  
\begin{eqnarray} \label{eq:localization}
& & \langle  \gamma(z') \prod_{\nu=1}^3 \Phi^{j_A,w_A}
 (x_\nu|\psi_\nu;z_\nu,\bar z_\nu)\, \prod_{\nu=4}^m \int d^2 z \Phi^{j_A,w_A}
 (x_\nu|\psi_\nu; z,\bar z) \prod_{i=1}^{R} \int d^2v \Phi^{\frac{1}{2b^2},-1}
 (v,\bar v)\rangle  \\[2mm]
 & & \hspace*{4cm} = \sum_\Gamma \Gamma(z') \langle \prod_{\nu=1}^m \Phi^{j_A,w_A}
 (x_A|\psi_A; z^\Gamma_A,\bar z^\Gamma_A)
 \prod_{i=1}^R \Phi^{\frac{1}{2b^2},-1} (v^\Gamma_i,\bar v^\Gamma_i)\rangle
\nonumber
\end{eqnarray}
provided that $R$ and the winding numbers are related by
\begin{equation} \label{eq:balance}
R = \frac12 \sum_{A=1}^m (w_A-1) + 1\ \quad \textit{ and } \quad w_A > 1 \ .
\end{equation}
Note that the first three fields are inserted at $(z_\nu, \nu = 1,2,3) = (0,1,\infty) $.
As long as the condition \eqref{eq:balance} is satisfied, we sum over covering maps 
$\Gamma$ which are discrete and the positions of all fields are localized at specific points on 
the worldsheet. Note that insertions of (integrated) vertex operators with $w_A = 1$ do not
contribute to the left hand side of the balancing condition eq.\ \eqref{eq:balance}. On the 
other hand they force $\Gamma$ to assume the value $x_A$ at the insertion point of the field. 
So, if we allow for insertions with $w_A =1$ the integration over their insertion points
is replaced by a discrete sum over the positions $u$ at which $\Gamma(u) = x_A$. 

We will not discuss the evidence or derivation of the localization formula \eqref{eq:localization} 
any further but simply refer to the original literature on the subject, see in particular 
\cite{Eberhardt:2019ywk}.\footnote{A path integral argument for the localization using 
the Kac-Wakimoto Lagrangian $S_0+S^-$ is about to appear in an upcoming paper by Bob 
Knighton, Sean Seet and Vit Sriprachyakul \cite{Knighton:2023mhq}.} Once we accept the 
validity of eq.\ \eqref{eq:localization} we can now explain our central formula 
\eqref{amplSL2R}. We will do so in two steps, starting with the case in which only 
the interaction term $S^-$ contributes. 
\medskip 
 
\paragraph{Analysis of $R_0$ insertions of $S^-$.} The string amplitude on the right 
hand side of equation \eqref{amplSL2R} may be computed by expanding the exponentials of 
the two interaction terms that appear in the theory \eqref{eq:S}. If we have less than 
$R_0$ insertions of $S_-$ the amplitude vanishes. Indeed, we note that the behavior of 
$\gamma$ near our $n$ vertex operators is given by eq.\ \eqref{eq:gammaPhiw} while the 
integrands of $S^-$ create first order poles in $\gamma$. If the number of such poles 
is too small, there simply is no covering map $\Gamma$ satisfying eq.\ \eqref{Gammadef}.
So, the first terms that can actually contribute arise when we have $R_0$ insertions of 
the first interaction term. In this case, there exists a discrete set of covering maps 
$\Gamma$ with $R_0$ poles or, put differently, a discrete set of branched coverings of 
genus $g=0$ with $R_0$ sheets. Hence the amplitude receives a discrete set of 
contributions that are in one-to-one correspondence with the genus $g=0$ contributions 
to the free symmetric product orbifold, 
\begin{eqnarray}
 & & \langle \prod_{\nu=1}^n \int d^2z \, \Phi^{j_\nu,w_\nu}
 (x_\nu|\psi_\nu;z) \rangle_{\mu} = \langle \prod_{\nu=1}^n
 \Phi^{j_\nu, w_\nu} (x_\nu|\psi_\nu; z_\nu)  \prod_{j=1}^{R_0} 
 \Phi^{\frac{1}{2b^2},-1}_{\frac{k}{2},\frac{k}{2}}(v_j) \rangle^{\Sigma_0} \ .
\nonumber
\end{eqnarray}
Note that the localization property of the path integral over $\gamma$ forces all the
worldsheet fields to be inserted at the zeroes and poles of $\partial \Gamma$. Comparison 
with formula \eqref{corrSL2R} shows that we have now almost reached the desired result, 
except that the vertex operators $\Phi_\nu$ still depend on the deformation parameter 
$x_\nu$ while those on the right hand side of eq.\ \eqref{corrSL2R} do not. But the 
difference is easy to remove using the Ward identities of the zero models $J^3_0$ and 
$\bar J^3_0$. These generate rescalings of the parameters $x_\nu$ and the associated 
Ward identities read
\begin{eqnarray} \nonumber
& & \langle \prod_{\nu=1}^n
 \Phi^{j_\nu, w_\nu} (u x_\nu|\psi_\nu; z_\nu) \ 
 \prod_{j=1}^{R_0}  \Phi^{\frac{1}{2b^2},-1}_{k/2,k/2}(v_j) \rangle^{\Sigma_0} =
 \\[2mm] \nonumber
& &  \quad \quad =  u^{- \sum_\nu (m_\nu + k w_\nu/2)} \bar u^{- \sum_\nu
(\bar m_\nu + k w_\nu/2)}
\langle \prod_{\nu=1}^n
 \Phi^{j_\nu, w_\nu} (x_\nu|\psi_\nu;z_\nu) 
 \prod_{j=1}^{R_0} \Phi^{\frac{1}{2b^2},-1}_{\frac{k}{2},\frac{k}{2}}(v_j)
 \rangle^{\Sigma_0}\ .
\end{eqnarray}
One might naively expect that we can use this simple behaviour under rescaling of the variables 
$x_\nu$ to send all of them to zero. But this turns out to be a bit too quick. As long as at least 
one of the variables $x_\nu$ is nonzero, it determines the scale for the overall prefactor $q$ 
that multiplies the derivative of the covering map $\Gamma$, see eq.\ \eqref{partialGamma}. This 
ceases to be the case when all $x_\nu$ vanish. In this sense, the limit $u\rightarrow 0$ is not 
a smooth limit. 

Since we cannot recover the prefactor $q$ of the covering map after we have sent $u$ to zero, we should 
make sure that we fix it before taking the limit. This is what our definition \eqref{eq:restrictedcorrelation}
of restricted correlations functions was designed for. A short analysis along these lines shows that there 
indeed exists some parameter $q' = \eta q$ such that 
\begin{eqnarray} \nonumber
\langle \prod_{\nu=1}^n
 \Phi^{j_\nu, w_\nu} (x_\nu|\psi_\nu; z_\nu) \ 
 \prod_{j=1}^{R_0} \Phi^{\frac{1}{2b^2},-1}_{\frac{k}{2},\frac{k}{2}}(v_j) \rangle^{\Sigma_0} & = & 
\langle \prod_{\nu=1}^n
 \Phi^{j_\nu, w_\nu} (\psi_\nu;z_\nu) \ \prod_{j=1}^{R_0}
 \Phi^{\frac{1}{2b^2},-1}_{\frac{k}{2},\frac{k}{2}}(v_j) \rangle^{\Sigma_0}_{q'}\  \\[2mm] 
 & & \hspace*{-2cm} \simeq \langle \prod_{\nu=1}^n \Phi^{j_\nu, w_\nu} (\psi_\nu;z_\nu) \ \prod_{j=1}^{R_0}
 \Phi^{\frac{1}{2b^2},-1}_{\frac{k}{2},\frac{k}{2}}(v_j) \rangle^{\Sigma_0}_{q} \ .
\end{eqnarray}
To go to the second line we have used the scaling behavior \eqref{eq:rescaling} of the reduced correlator 
and we dropped the factor $\Lambda(\eta)$ in front of the correlator which is why we wrote $\simeq$ 
rather than an equal sign. Note that the explicit $x_\nu$ dependence has now disappeared from the field 
insertions on the right hand side. Nevertheless, the correlator certainly continues to depend on $x_\nu$ 
through the dependence of the insertion points $z_\nu$ and $v_j$ on $x_\nu$. We recall that the latter 
are branch points and poles of the covering map $\Gamma$ whose defining property \eqref{Gammadef} does 
depend on $x_\nu$. Thereby we have now expressed the first non-vanishing term of the original string 
amplitude \eqref{amplSL2R} through the same restricted correlator that appears on the right hand side 
of our formula \eqref{corrSL2R} for the correlation function in the symmetric product orbifold. 

So far we have not considered the effect of inserting the second interaction term $S^+$. So, let us 
now imagine that we bring down some $S^+$ insertions from the exponential while keeping the number 
of $S^-$ insertions fixed at the minimal value $R_0$. As we have saw in eq.\ \eqref{eq:triviality}, 
insertions of $S^+$ into such a correlator are actually trivial. Hence the terms with $R_0$ 
insertions of $S^-$ reproduce the correlators of the free symmetric product orbifold. This is quite 
nice already, but things become truly remarkable once we increase the number of $S^-$ insertions 
beyond the smallest value $R_0$.
\medskip 

\paragraph{More than $R_0$ insertions of $S^-$.}
To see this, let us first consider the case in which we insert just one addition screening charge $S^-$, 
i.e. we have $R = R_0+1$.\footnote{Here we assume that $R_0$ is integer. If, on the other hand, $R_0$ 
was half-integer, the terms without any $S^+$ insertions would vanish and the first non-vanishing terms
would arise from $R= R_0 + 1/2$ insertions of $S^-$ and two insertions of $S^+$.}  The additional 
insertion of $S^-$ creates one more pole in $\gamma$ or, equivalently, an additional second order 
pole with vanishing residue in $\partial \gamma$. Consequently, $\partial \gamma$ is now
allowed to have two additional zeroes beyond those that are associated with the $n$
vertex operators $\Phi_\nu$. Since our second interaction term has winding number
$w=1$, it creates a single zero in $\gamma$, see eq.\ \eqref{gammaQ}, which is of
course no longer visible in $\partial \gamma$. So, at first sight it might seem
that we cannot get the desired zeroes from insertions of $S^+$. But this is not
true. Let us assume that we brought down four interaction terms of the second
kind. Since we integrate over the insertion points of these interaction terms,
our integration includes regions in which two of the vertex operators with $w=1$
come close to each other. If that is the case, they generate an operator with
winding number $w=2$,
\begin{equation} \label{eq:QpQpope}
\Phi^{\frac{1}{2b^2},1}_{-\frac{k}{2},-\frac{k}{2}}(u_1,\bar u_1)
\, \Phi^{\frac{1}{2b^2},1}_{-\frac{k}{2},-\frac{k}{2}}(u_2,\bar u_2)
\sim |u_1-u_2|^2 \Phi^{\frac{1}{b^2},2}_{-k,-k}(u_2,\bar u_2) + \dots \, .
\end{equation}
The vertex operator with winding number $w=2$ that appears in the leading
term of the operator product expansion creates a zero of order two in $\gamma$
and hence a single zero in $\partial\gamma$. Hence, if the four screening charges
we have brought down from the second interaction term form two such pairs,
$\partial \gamma$ will have two additional zeroes, just as desired, and the
correlation function is no longer zero. In other words, with $R=R_0+1$ insertions
of the screening charge $S^-$, our amplitude receives nontrivial contributions
from the boundary of the integration region of four $S^+$ insertions where these
collide pairwise. Since the leading term in the operator product \eqref{eq:QpQpope}
contains the very same operator that was required in our formula \eqref{corrSL2R}
in order to reproduce insertions of the perturbing operator of the symmetric product 
orbifold, we conclude that these boundary contributions are indeed given by the right 
hand side of our formula \eqref{corrSL2R}. The only difference is that the vertex 
operators at $z_\nu$ are deformed by $x_\nu$. 

The arguments outlined in the previous paragraph obviously rely on the localization 
\eqref{eq:localization}. The latter involves operators with positive winding that are 
deformed. For the vertex operator insertions with winding number $w=1$ that arise from 
the second interaction term $S^+$, the deformation can be introduced through the 
following identity 
\begin{align} \label{secondsov}
\Phi^{\frac{1}{2b^2},1}_{-\frac{k}{2},-\frac{k}{2}} (u)  \propto
\int \frac{d ^2 \xi }{|\xi|^2} e^{- \xi \beta_0 - \bar \xi \bar \beta_0} 
\Phi^{\frac{1}{2b^2},1}_{-\frac{k}{2},-\frac{k}{2}} (u)  \, .
\end{align}
If we insert the our interaction term at some point $u$ on the worldsheet, there 
exists some value of the deformation parameter $\xi$ for which this insertion is 
non-zero, namely the value $\xi = \Gamma(u)$ the branching function $\Gamma$ 
assumes  at $u$. Since the right hand side of the identity \eqref{secondsov} 
involves an integration over all $\xi$, we can pick up some non-zero contribution 
for any value of $u$. Consequently, the integration over the insertion point of 
the interaction vertex remains intact while the integration over the associated 
deformation parameters localizes. By this mechanism, all operators with positive 
winding number are indeed deformed. 

The pairwise collision mechanism we have just described allowed four insertions of
$S^+$ to give non-trivial contribution to the amplitudes with $R_0+1$ insertions of
$S^-$. Additional insertions of $S^+$, on the other hand, give trivial contributions
again. Indeed, their collisions would produce additional double zeroes of $\gamma$ and
hence single zeroes in $\partial\gamma$. But this is not allowed as long as we do not
increase the number of poles (or change the genus of the underlying surface). The
extension to higher number of $S^-$ insertions is now straightforward. Each insertion
of $S^-$ creates two new poles in $\partial \gamma$ and hence, for fixed genus $g$,
also two new zeroes. We can think of these as holes that can now trap a pair of
$S^+$ insertions. Following the discussion in the previous paragraph, the resulting
string amplitude with $R= R_0 + s/2$ is given by
\begin{eqnarray}
 & & \langle \prod_{\nu=1}^n \int d^2z \, \Phi^{j_\nu,w_\nu}
 (x_\nu|\psi_\nu,z,\bar z) \rangle_{\mu} \simeq \\[2mm]
& & \hspace*{2cm} \simeq  \sum_{s=0}^\infty \frac{\mu^{2s}}{s!}
\int \prod_{a=1}^{s}  d ^2 u_a \langle \prod_{\nu=1}^n
 \Phi^{j_\nu, w_\nu} (\psi_\nu; z_\nu)  \prod_{a=1}^{s}
 \Phi^{\frac{1}{b^2},2}_{-k,-k} (u_a)  \!\!
 \prod_{j=1}^{R_0+s/2}\!\! \Phi^{\frac{1}{2b^2},-1}_{\frac{k}{2},\frac{k}{2}}(v_j) \rangle^{\Sigma_0}_q \ .
\nonumber
\end{eqnarray}
Note that the localization property of the path integral over $\gamma$ forces all the
worldsheet fields to be inserted at the zeroes and poles of $\partial \Gamma$. Hence,
only the integration over the insertion points $u_a$ remains in this expression since 
these are not fixed by the constraints on the branching function. Following the same 
arguments we outlines in the previous paragraph where we discussed the case of $R_0$ 
insertions of $S^-$, we have removed the deformation parameters $x_\nu$ of the vertex 
operators on the right hand side by some global rescaling. In this step, the expectation 
value $\langle \cdot \rangle$ is replaced by the restricted one $\langle \cdot \rangle_q$, 
up to some factor of the form \eqref{eq:Lambda}. Since we have not been able to compute 
the value of the relevant $\eta$, we dropped this prefactor and wrote $\simeq$ instead 
of an equality. Thereby we have now expressed the original string amplitude 
\eqref{amplSL2R} through the same restricted correlator that appears on the right hand 
side of our formula \eqref{corrSL2R} for the correlation function in the deformed symmetric 
orbifold. In conclusion, we have reached our main goal to show that the string amplitudes 
in the bosonic string theory \eqref{eq:S} reproduce the correlation functions in the 
perturbed symmetric product orbifold, see eq.\ \eqref{amplSL2R}.

\section{Generalization to Higher Genus Surfaces}
\label{section.6}

In the previous sections we have shown that the leading planar contributions to the
correlation functions of the perturbed symmetric orbifold theory are captured by the
genus $g=0$ string amplitudes of the bosonic string theory \eqref{eq:S}. Our goal now
is to extend this analysis to the non-planar corrections. We will indeed show that
these are reproduced by higher genus amplitudes in the same bosonic string theory.
In the first subsection we shall describe contributions to the correlation functions
in the symmetric product orbifold that arise from branched coverings of genus $g$. The
main goal here is to extend formula \eqref{corrdef} to $g > 0$. Then we turn our
attention to the WZNW model and establish the higher genus analogue of our formula
\eqref{corrSL2R} in the second subsection. The uplift to pure bosonic string theory
will then work in precisely the same as in the previous section.

\subsection{Symmetric orbifolds beyond the planar limit}

As argued in section \ref{section.2}, the correlation functions of twisted sector fields
in the symmetric product orbifold $\mathcal{M}^N/S_N$ on a sphere can be computed as correlation 
functions of single CFT $\mathcal{M}$ on a Riemann surface defined by the covering map $x = 
\Gamma (z)$. Here $\Gamma(z)$ is required to satisfy the condition \eqref{Gammadef}. In 
section \ref{section.2}, we assumed that the covering map defines a Riemann surface 
$\Sigma = \Sigma_0$ of genus zero. In this subsection, we generalize the analysis to 
the case where the covering map defines a Riemann surface $\Sigma = \Sigma_g$ of 
generic genus $g \geq 1$. Recall that the genus $g$ is related to the number $R$ of 
closed solid loops of the diagram through eq.\ \eqref{eq:gRrel}. 

We start by reviewing some useful facts on higher genus Riemann surface. Our exposition 
follows \cite{Verlinde:1986kw,Hikida:2007tq,Hikida:2008pe}, see also \cite{Fay:703829,
Mumford:109230,AlvarezGaume:1986es}. On the Riemann surface $\Sigma_g$, we first introduce 
a complex structure so that we have a notion of holomorphicity. It is well known that a 
complex surface of genus $g$ admits $g$ independent holomorphic one-forms $\omega_a$. 
We choose a basis in the space of holomorphic one-forms as 
\begin{align}
\oint_{\alpha_a} \omega_b = \delta_{a,b} \, , \quad \oint_{\beta_a} \omega_b = \tau_{ab}
\, . \label{cycles}
\end{align}
Here $\alpha_a,\beta_a, a = 1, \dots, g,$ denote a basis of holomorphic cycles on the 
surface and $\tau_{ab}$ is the period matrix of $\Sigma_g$. In order to construct some 
coordinates on $\Sigma$, we utilize the Abel map
\begin{align}
z_a = \int_{z_0}^z \omega_a \, \label{Abel}
\end{align}
where $z_0$ is an arbitrary reference point in $\Sigma_g$. We shall need a number of 
basic function on our surface. All of these are constructed in one way or another form 
the theta functions
\begin{align}
\theta_\delta (z | \tau) = \sum_{n \in \mathbb{Z}^g} \exp\left\{  i \pi
[(n+ \delta_1)^a \tau_{ab} (n+ \delta_1)^b + 2 ( n + \delta_1)^a (z+ \delta_2)_a ]\right\} \, ,
\end{align}
where $\delta_a  = (\delta_{1a} , \delta_{2a})$ with $ \delta_{1a},\delta_{2a} = 0,1/2$
defines the spin structure along the homology cycles $\alpha_a,\beta_a$. The theta function
is actually not periodic along the holomorphic cycles. Instead, under shifts of the form 
$m+\tau n$, it behaves as 
\begin{align}
\theta_\delta (z  + m + \tau n|\tau) = \exp [ - i \pi (n^a \tau_{ab} n^b + 2 n^a z_a)]
\exp [ 2 \pi i (\delta^a_1 m_a - \delta_2^a n_a ) ] \theta_\delta (z|\tau) \, .
\end{align}
According to the Riemann vanishing theorem, the theta function vanishes at a point $z$ if
and only if one can find $g-1$ points $p_i$ $(i=1,\ldots,g-1)$ such that $z$ can be written 
as 
\begin{align}
z = \Delta - \sum_{i=1}^{g-1} p_i \ . \label{Riemann}
\end{align}
Here, $\Delta$ denotes the Riemann class. This concludes our description of background
material on higher genus Riemann surfaces and it suffices to construct the relevant
covering maps.

The covering map $\Gamma:\Sigma_g \mapsto S^2$ is defined such as to satisfy the conditions 
\eqref{Gammadef} near the branch points at $z = z_A$ of the surface $\Sigma_g$. As in the 
case of genus zero, it is convenient to consider the derivative $\partial \Gamma$. The 
meromorphic one-form $\partial \Gamma$ has a zero of order $w_A -1$ at $z = z_A$, thus the 
total order of zeros is $2 R^{(g)} -2 + 2 g$, where $R^{(g)}$ is defined by \eqref{eq:Rg}. 
The covering map $\Gamma$ has $R^{(g)}$ poles of the first order, whose locations are 
denoted by $v_j$, as before, and hence the derivative $\partial \Gamma$ should have 
$R^{(g)}$ poles of the second order at $z =v_j$. From this information of zeros and 
poles, the derivative $\partial \Gamma$ can therefore be written as 
\begin{align} \label{eq:Gammagdef} 
    \partial \Gamma (z) = \frac{q \prod^m_{A=1} E (z , z_A)^{w_A -1}
    \sigma (z)^2}{\prod_{j =1}^R E(z ,v_j)^2} \, .
\end{align}
The function $E(z, w)$ that appears in the numerator and denominator of the formula for 
$\partial \Gamma$ is a prime form which is defined as
\begin{align}\label{primedef}
E (z,w) = \frac{\theta_\delta (\int^z_w \omega | \tau )}{h_\delta (z) 
h_\delta (w)} \, , \quad  \textit{where}  \quad 
h_\delta (z)^2 = \sum_{a=1}^g \partial_a \theta_\delta (0|\tau) \omega_a (z) \, .
\end{align}
In this expression the spin structure $\delta$ is assumed to be odd. The prime form has
weight $(-1/2,0)$ both for $z$ and $w$ and has a zero of the first order at $z=w$, i.e.\ it
behaves as $E(z ,w) \sim z -w$. In addition, the prime form $E$ is periodic along
$\alpha_a$ cycles, but it is not so along $\beta_a$ cycles. More precisely, upon 
shifts of $z$ by $\tau_a$ the prime form behaves as
\begin{align} \label{eq:Eshift}
E (z + \tau_a , w) = - \exp \left( - i \pi \tau_{aa} - 2 \pi i \int^w_z \omega_a  \right) E(z,w) \, .
\end{align}
In order to turn $\partial \Gamma$ into a proper (double periodic) one-form on the surface, 
we put the factor $\sigma(z)^2$ in the numerator on the right hand side of eq. 
\eqref{eq:Gammagdef}. The object $\sigma(z)$ is a $g/2$-form that is defined by
\begin{align}
\ln | \sigma (z) |^2  =  \frac{1}{16 \pi } \int d^2 w \sqrt{g (w)}
\mathcal{R} (w) \ln |E (z,w)|^2  \ .  \label{sigmadef}
\end{align}
Upon shifts of $z$ by $\tau_a$, the $g/2$-form $\sigma$ behaves as
\begin{align} \label{eq:sigmashift} 
\sigma (z + \tau_a , w) =  \exp \left( - \pi i (g-1) \tau_{aa}  +
2 \pi i \int^\Delta_{(g-1)z } \omega_a \right) \sigma (z  , w)  \, .
\end{align}
With the help of the shift properties \eqref{eq:Eshift} and \eqref{eq:sigmashift}, we can 
now compute the behaviour of  $\partial \Gamma(z)$ under shifts of $z$. It is then easy to 
see that $\partial \Gamma$ is a single-valued one-form on $\Sigma_g$, provided that the 
parameters $v_j$ that appear in the denominator of \eqref{eq:Gammagdef} satisfy the 
condition
\begin{align}
 \sum_{\nu=1}^n (w_\nu -1) \int_{z_0}^{z_\nu} \omega_a - 2
 \sum_{j =1}^R \int_{z_0}^{v_j} \omega_a - 2 \int_{(g-1)z_0}^\Delta \omega_a = 0 \, .
\end{align}
After these preparations, we can write down the expression of correlation function of twisted 
sector fields in the symmetric product orbifold. Utilizing the result obtained in \cite{Hikida:2020kil}, 
the genus $g$ expression can be found to take the form  
\begin{align} \label{freecorrg}
 &\langle \prod_{A=1}^{m} V^{(w_A)}_{\alpha_A}(\psi_A;x_A) \rangle^{S^2} =
  e^{\frac{3k}{4}U_g}\sum_{\Gamma} \prod_{A =1}^{m} 
(\tilde a_A^\Gamma)^{-h_{\alpha_A}^{(w_A)} + \frac{k}{4} (w_A -1)} (\bar{\tilde a}_A^\Gamma)^{-\bar h_{\alpha_A}^{(w_A)} + \frac{k}{4} (w_A -1)} \\&  \times \prod_{A =1}^{m}  \sigma (z_A)^{- 2 h^{(w_A)}_{\alpha_A}(\psi_A)
 +  k (w_A -1)} \bar \sigma (z_A)^{- 2 \bar h^{(w_A)}_{\alpha_A}(\psi_A)
 +  k (w_A -1)}
 \prod_{j =1}^R |\xi^\Gamma_j|^{- k} |\sigma (v_j)|^{- 4 k} \langle  \prod_{A=1}^{m}
 V_{\alpha_A}(\psi_A;z_A) \rangle^{\Sigma_g} \, . \nonumber
 \end{align}
In order to write this formula in a reasonably compact form we have introduced functions
\begin{align} \label{axig}
\tilde a_A^\Gamma = \frac{w_A a_A^\Gamma}{\sigma (z_A)^2}
= \frac{q\, \prod_{B \neq A } E(z_A ,z_B)^{w_B -1}}{\prod_{j=1}^R
E(z_A , v_j)^2} \, , \quad
\xi_j ^\Gamma = \frac{q\, \prod_{A=1}^m E (v_j , z_A)^{w_A -1}}{\prod_{i \neq j }
E(v_j , v_{i})^2}
\end{align}
and
\begin{align}
U_g = \frac{1}{192 \pi^2} \int d^2z d^2 w \sqrt{g (z)} \mathcal{R }(z) \sqrt{g(w)}
\mathcal{R} (w) \ln |E (z,w)|^2 \, . \label{Ug}
\end{align}
These formulas are the obvious extensions of the formulas \eqref{freecorr} and 
\eqref{axi} to surfaces of genus $g > 0$. 
 
We are interested in computing correlation function of the form \eqref{corrdef}. In order 
to obtain the contributions to these correlators that are associated with surfaces of genus 
$g> 0$, we may set $m = n+s$ and $w_A=2, \alpha_A = 1/2b, \psi_A = |0\rangle$ and
$x_A = y_{A-n}$ for all $A > n$ in eq.\ \eqref{freecorrg}. For coverings of genus $g$, the  
momentum conservation condition \eqref{eq:alphaconstraint} is replaced by
\begin{align} \label{eq:alphaconstraintg}
 \sum_{\nu=1}^n \alpha_n + \frac{s}{2b} = Q_\varphi (1 -g) \, .
\end{align}
From eq.\ \eqref{axig} with $\tilde b^\Gamma_a = \tilde a^\Gamma_{n+a}$ for $a =1,2,\ldots, s$, we
find 
\begin{align} \label{abxig}
\begin{aligned}
& \tilde a_\nu^\Gamma = \frac{w_\nu a_\nu^\Gamma}{\sigma (z_\nu)^2} =
\frac{q\, \prod_{\mu \neq \nu}^{n} E(z_\nu , z_{\mu})^{w_{\mu} -1} \prod_{a=1}^s
E(z_\nu ,u_a)}{ \prod_{j=1}^{R_0+s/2} E (z_\nu , v_j)^2 } \, , \\[2mm]
& \tilde b^\Gamma_a =\frac{2 b^\Gamma_a}{\sigma (u_a)^2}
 = \frac{q\, \prod_{\nu=1}^n E (u_a , z_{\nu})^{w_\nu -1} \prod_{b \neq a}^{s}
 E (u_{a} , u_{b})}{ \prod_{j=1}^{R_0+s/2} E (u_a ,v_j)^2 } \, , \\[2mm]
&\xi_j^\Gamma = \frac{q\, \prod_{\nu=1}^n E (v_j , z_\nu)^{w_\nu -1}\prod_{a=1}^s
E (v_j , u_a)}{\prod_{i \neq j}^{R_0 + s/2} E(v_j , v_i)^2} \, .
\end{aligned}
\end{align}
With these definitions, the contributions of genus $g$ coverings to the correlation function
\eqref{corrdef} can be expressed as
\begin{align}
 &\langle \prod_{\nu=1}^n V^{(w_\nu)}_{\alpha_\nu} (\psi_\nu;x_\nu) \rangle ^{S^2}_\lambda
  = \sum_{\Gamma_g} \sum_{s=1}^\infty \frac{\lambda^s}{s!} e^{\frac{3}{4}k U_g}\int \prod_{a=1}^s d^2 u_a
  \prod_{a=1}^s |\tilde b_a^\Gamma|^{ \frac{k}{2}}
 |\sigma (u_a)|^{2k} \prod_{j=1}^{R_0+s/2} |\xi_j|^{-k} |\sigma (v_j)|^{-4k}
 \nonumber
  \\[2mm]
 & \hspace*{0cm}  \times 
  \prod_{\nu =1}^{n} 
(\tilde a_\nu^\Gamma)^{-h_{\alpha_\nu}^{(w_\nu)} + \frac{k}{4} (w_\nu -1)} (\bar{\tilde a}_\nu^\Gamma)^{-\bar h_{\alpha_\nu}^{(w_\nu)} + \frac{k}{4} (w_\nu -1)}  \sigma (z_\nu)^{- 2 h^{(w_\nu)}_{\alpha_\nu}(\psi_\nu)
 +  k (w_\nu -1)} \bar \sigma (z_\nu)^{- 2 \bar h^{(w_\nu)}_{\alpha_\nu}(\psi_\nu)
 +  k (w_\nu -1)} \nonumber \\
 & \hspace*{0cm}  \times 
 \langle \prod_{\nu=1}^n V_{\alpha_\nu} (\psi_\nu ; z_\nu)
 \prod_{a=1}^s V_{\frac{1}{2b}} (u_a) \rangle ^{\Sigma_g}\, .   \label{deformedcorrg}
\end{align}
In writing this expression we have also changed the integration variables
from $y_a$ to $u_a$ using the Jacobian
\begin{equation}
d^2y_a = d^2 u_a |\tilde b^\Gamma_a|^2 |\sigma(u_a)|^4 \ .
\end{equation}
Formula \eqref{deformedcorrg} extends the previous formula
\eqref{deformedcorr} to branched coverings of arbitrary genus $g$. It is this
expression for nonplanar contributions to the correlation functions of the
perturbed symmetric product orbifold that we now need to embed into the
WZNW model on the surface $\Sigma_g$.

\subsection{Embedding into the WZNW model for higher genus}
 
In the case of genus zero, the essential ingredient of our construction was the proof of the 
formula \eqref{corrSL2R}. The higher genus extensions that we are going to establish in this 
subsection takes the form 
\begin{align} \label{corrSL2Rg}
 \langle \prod_{\nu=1}^n V^{(w_\nu)}_{\alpha_\nu} (\psi_\nu;x_\nu)
 \rangle^{S^2}_\lambda & := \sum_{\Gamma_g} \sum_{s=1}^\infty \frac{\lambda^s}{s!}
 \int \prod_{a=1}^{s}  d ^2 u_a \times \\[2mm] 
 & \hspace*{2cm} \times \langle \prod_{\nu=1}^n
 \Phi^{j_\nu, w_\nu} (\psi_\nu;z_\nu)  \prod_{a=1}^{s}
 \Phi^{\frac{1}{b^2},2}_{-k,-k} (u_a)  \!\!
 \prod_{j=1}^{R_0+s/2}\!\! \Phi^{\frac{1}{2b^2},-1}_{k/2,k/2}(v_j) \rangle^{\Sigma_g}_q \, .
\nonumber
\end{align}
Our notations are the same as in eq.\ \eqref{corrSL2R}. In particular, the vertex operators of 
the Kac-Wakimoto free field theory on the right hand side are inserted at the zeroes and poles of 
$\partial \Gamma$, see eq.\ \eqref{eq:Gammagdef}. The definition of the restricted correlation 
functions is similar the one we have in eq.\ \eqref{eq:restrictedcorrelation}, but with $\hat u$ 
defined, e.g.\ by equating the two expressions \eqref{betag} and \eqref{mu2upg} for $\hat \beta$ 
below. An explicit formula for $\hat u$ is not as easy to write down in this case. Fortunately, 
we do not really need it anyway. In order to establish equation \eqref{corrSL2Rg}, we start form 
the correlation function of the Kac-Wakimoto free field theory that appears in the integrand on 
the right hand side, i.e. 
\begin{equation} \label{Cg}
C^{(n;s)}_{g}\left({\genfrac{}{}{0pt}{}{j_\nu,w_\nu} {m_\nu,\bar m_\nu}} \middle| \,
{\genfrac{}{}{0pt}{}{z_\nu,u_a}{v_j, q}} \right) :=
\langle \prod_{\nu=1}^n
 \Phi^{j_\nu, w_\nu}_{m_\nu,\bar m_\nu} (z_\nu)  \prod_{a=1}^{s}
 \Phi^{\frac{1}{b^2},2}_{-k,-k} (u_a)  \!\! \prod_{j=1}^{R_0+s/2}\!\!
 \Phi^{\frac{1}{2b^2},-1}_{\frac{k}{2},\frac{k}{2}}(v_j) \rangle^{\Sigma_g}_q \, ,
\end{equation}
and show that it reproduces the integrand in the right hand side of eq.\ \eqref{deformedcorrg}
for the correlation function in the symmetric product orbifold. Here, the formula \eqref{eq:Rg} 
leads to
$$ R^{(g)} = \frac12 \sum_{A=1}^{n+s} (w_A-1) + 1 - g = \frac12 \sum_{\nu=1}^n (w_\nu-1) +
\frac{s}{2} +1 - g \equiv R^{(g)}_0 + \frac{s}{2}$$
with $R=R^{(g)}$ and $R_0 = R^{(g)}_0$. The charge conservation of the linear dilaton $\phi$ 
requires that
\begin{align}
 \sum_{\nu=1}^n j_\nu - \left(\sum_{\nu=1}^n w_\nu - 2 R \right) \frac{1}{2b^2} = 1- g \, ,
\end{align}
which coincides with the momentum conservation condition \eqref{eq:alphaconstraintg} if we 
apply the substitution \eqref{js2}.

We evaluate the quantity \eqref{Cg} by following the same strategy as in subsection \ref{section.4.2}.
Namely, in a first step we utilize the parafermion representation to collect all the spectral flow in 
a single insertion point. For surfaces of higher genus, the parafermionic representation requires a 
little more thought since there is a subtlety associated with phases (twists) which arise when going 
along non-trivial cycles of Riemann surfaces. As reviewed in appendix A of \cite{Hikida:2008pe} (see 
also \cite{Gawedzki:1988nj,Gawedzki:1991yu,Martinec:1991ea,Dijkgraaf:1991ba}), the correlation 
functions of the parafermionic primary fields are given by
\begin{align} \label{FP}
\langle \prod_{I=1}^M \Psi^{j_I}_{m_I , \bar m_I} (z_I) \rangle = \Delta_\text{FP} 
\int \mathcal{D}g \mathcal{D}\chi \prod_{a=1}^g d^2 \Lambda_a e^{-S^\text{cig} [g,\chi]_\Lambda} 
\prod_{I=1}^M\Psi^{j_I}_{m_I , \bar m_I} (z_I ) \, .
\end{align}
Here $\Delta_\text{FP}$ is the Fateev-Popov determinant and $S^\text{cig}[g,\chi]_\Lambda$ is the 
sum of the action of the WZNW model with $g \in \textit{SL}(2)$ and a free boson $\chi$. Moreover, 
the parameter $\Lambda_a$ represents the twist along $\beta_a$ cycle as
\begin{align} \label{twists}
\begin{aligned}
&\beta  (z_a + \tau_{ab} n^b + m_a | \tau ) = e^{2 \pi i n^b \Lambda_b} \beta (z_a | \tau) \, ,  \\[2mm]
&\gamma   (z_a + \tau_{ab} n^b + m_a | \tau ) = e^{- 2 \pi i n^b \Lambda_b} \gamma (z_a | \tau) \, ,  \\[2mm]
&\phi (z_a + \tau_{ab} n^b + m_a | \tau) = \phi (z_a | \tau) + \frac{2 \pi n^c \text{Im} \Lambda_c}{b} \, ,
\end{aligned}
\end{align}
where $\tau_{ab}$ is the period matrix of the surface $\Sigma$ that we introduced previously in eq.\ \eqref{cycles}.

From the definition of correlation functions in the parafermionic theory \eqref{FP}, we can obtain the relation,
\begin{align}\label{eq:paraAC}
\langle \, \prod_{I=1}^M \Psi^{j_I}_{ m_I ,  \bar m_I} (z_I)  \,  \rangle
= |\det {}' \partial|^2 \int \prod_{a=1}^g d^2 \Lambda_a A_M^\Lambda |B_M|^{2}   =   
|\det {}' \partial|^2 \int \prod_{a=1}^g d^2 \Lambda_a  C_M^\Lambda |D_M|^{2}  \, ,
\end{align}
by utilizing the fact that the parafermionic fields do not depend on the winding number $w$ as in 
eq.\ \eqref{cosetlang}. Furthermore, $ |\det ' \partial|^2$ arises from the integration over Fadeev-Popov 
ghosts, with the prime indicating that the zero mode contribution is removed. The correlators $A_M^\Lambda$ 
and $C_M^\Lambda$ that appear in the integrands on the right hand side of eq.\ \eqref{eq:paraAC} are defined 
by
\begin{align}
A_M^\Lambda =
\langle\,  \prod_{I=1}^M \Phi^{j_I , w_I}_{m_I , \bar m_I} (z_I)  \,  \rangle_{\Lambda,q} \, , \quad
C_M^\Lambda =
\,  \langle v^{(w)} (\xi)\prod_{I=1}^M \Phi^{j_\nu , 0}_{m_I , \bar m_I} (z_I) \, \rangle_{\Lambda,q} \, .\label{Cnpg}
\end{align}
They are evaluated with the \SL2R WZNW model in the presence of twists $\Lambda$ that were introduced in eqs. 
\eqref{twists}. The dependence on the twist parameters is indicated by the subscript $\Lambda$ on the 
correlators. This is not to be confused with the index $q$ we introduced before to denote the restricted 
correlation functions of the Kac-Wakimoto free field theory. We also introduces the integer $w = \sum_I 
w_I$ by summing over all the individual winding numbers, as in eq.\ \eqref{eq:wsumw}. The other correlators 
that appear in the integrands of eq.\ \eqref{eq:paraAC} are computed within the free bosonic theory as 
\begin{align}
B_M = \langle \prod_{I =1}^M e^{i\sqrt{\frac{2}{k}} (m_I + \frac{kw_I}{2}) \chi(z_I)} 
\rangle \, , \quad D_M =  \langle e^{i\sqrt{\frac{k}{2}} w \chi (\xi)}
\prod_{I =1}^M e^{i\sqrt{\frac{2}{k}} m_I \chi(z_I)}  \rangle \, .
\end{align}
In the case of genus zero, there is no non-trivial cycle and twist, thus the correlators of parafermionic 
primary fields can be written by the products of those of the WZNW model and the free boson theory. However,
in the case of higher genus, the relation between the two different correlation functions of the WZNW model 
is not so simple and twists along non-trivial $\beta$ cycles need to be introduced. We are interested in 
the correlation functions of the WZNW model of the form of $A^\Lambda_M$, without any twists. In our special 
type of correlation functions, the derivative $\partial \gamma (z)$ of the ghost field $\gamma$ can be 
identified with the derivative $\partial \Gamma(z)$ of the covering map. This implies that the integration 
over the twists $\Lambda_a$ is localized at $\Lambda_a = 0$. For this reason, we can ignore the additional 
complications that arise from non-trivial twists. 

Given eq.\ \eqref{eq:paraAC}, we can compute the desired quantity $A^\Lambda_M$ from the correlations 
function $C^\Lambda_M$ in the integrand on the right hand side. As explained in subsection 
\ref{section.4.2},  it is convenient to convert $\beta\gamma$ system to $\hat \beta \hat \gamma$ system  via the 
transformation \eqref{eq:hatbetagamma} in order to reflect the spectral flow. As in the previous discussion, 
this leads to the following new expression for the correlation function $C^\Lambda_M$, 
\begin{align}
C_M^\Lambda = q^{-\sum_\nu (m_\nu + \bar m_\nu + k w_\nu)}
\int \prod_{I=1}^{M} \frac{d^2 \mu_I}{|\mu_I|^2} \mu_I^{m_I} \bar \mu_I^{\bar m_I}
 \delta^{(2)}(\hat u-1)   \langle \hat v^{(w)} (\xi)
\prod_{I=1}^{M}  \hat \Phi^{j_I} (\mu_I | z_I)\, 
 \rangle \, .
\end{align}
Here, the fields $\hat \Phi$ and $\hat v$ are the same as 
in our definitions \eqref{mubasis} and \eqref{hatu}. Having inverted the spectral flow we can now proceed as in 
\cite{Hikida:2007tq,Hikida:2008pe}. Integrating over $\hat \gamma , \hat \beta $, we obtain 
\begin{align}
\hat \beta  (z) = \sum_{I =1}^M \frac{1}{\mu_I} \sigma_\Lambda (z , z_I) + \sum_{\sigma =1}^{g-1} 
\varpi_\sigma \omega^\Lambda_\sigma (z) \, . \label{betag}
\end{align}
In comparison to our earlier expression \eqref{mu2up} for surfaces of genus $g=0$ the expression now also 
includes a sum over the basis $\omega_\sigma^\Lambda(z)$ of $\Lambda$ twisted holomorphic differentials. The 
function $\sigma_\Lambda (z,w)$ that contains the contributions from the insertion points $z_I$ is given 
by 
\begin{align}
\sigma_\Lambda (z ,w) = \frac{(h_\delta (z))^2  \theta_\delta (\Lambda - \int^z_w \omega)}
{\theta_\delta (\int^z_w \omega) \theta_\delta (\Lambda)}
\end{align}
with odd spin structure $\delta$.
Following the same strategy as in the case of genus $g=0$, see in particular eq.\ \eqref{mu2up} we can 
express the right hand side of eq.\ \eqref{betag} in the form 
\begin{align}
\hat \beta  (z) =
\hat u \frac{E (z , \xi)^{w}  \sigma (z)^2}{\prod_{I=1}^{M }E(z , z_I) } \, . \label{mu2upg}
\end{align}
Here we set $w = M + 2 g -2$.
%since it is enough to consider the case for our purpose.
The parameters $\mu_I$ are related to the residues of the function $\hat \beta$ at the 
insertion points $z = z_I$ through 
\begin{align}
\frac{1}{\mu_I} = \hat  u \frac{E (z_I , \xi)^{w}  \sigma (z_I)^2}{\prod_{J \neq I }^{n  }E(z_J , z_I) } \, . \label{mug}
\end{align}
In order for eq.\ \eqref{mu2upg} to be satisfied, the following $w$ constraint equations need 
to be satisfied 
\begin{align}
f_{m,\xi} (\mu , \varpi , \Lambda)  := \sum_{I=1}^M \frac{1}{\mu_I} \sigma_\Lambda^{(m-1)} 
(z_I , \xi) + \sum_{\sigma =1}^{g-1} \varpi_\sigma \omega_\sigma^{\Lambda (m-1)} (\xi ) = 0
\end{align}
for $m=1,\dots, w$.
The Jacobian for the transformation from the variables $\mu_I, \varpi_\sigma$ and $\Lambda_a$ is now 
given by, see \cite{Hikida:2008pe}, 
\begin{align}
&\prod_{I=1}^{M} \frac{d^2 \mu_I}{|\mu_I|^2} \prod_{\sigma = 1}^{g-1} d^2 \varpi_\sigma \prod_{a=1}^g d^2 \Lambda_a
 \prod_{m=1}^{w} \delta^2 \left( f_{m,\xi} (\mu , \varpi ,\Lambda) \right) \frac{|\det ' \partial |^2}{|\det ' \nabla _\Lambda |^2 }   = \frac{d \hat u}{| \hat u |^{4 - 2g + 2 w}} \frac{\prod_{I < J}|E(z_I , z_J)|^2 }{ \prod_{I} |\sigma (z_I)|^2 |\sigma (\xi)|^{2w} }  \, .
\end{align}
Here $1/|\det ' \nabla _\Lambda |^2 $ comes from the integration over $\hat \beta \hat \gamma $ ghost system.
If we now shift the field $\phi(z, \bar z)$ as
\begin{align}
\varphi (z , \bar z) = \phi (z , \bar z)  - \frac{1}{2 b} \left( w  \ln |E (z , \xi)|^2 - \sum_{I =1}^{M } |E (z , z_I)| ^2 + 2 \ln |\sigma (z)|   -  \ln | \rho(z)|^2 \right) 
\end{align}
and perform the integration over the variables $\mu_I , \varpi_\sigma , \Lambda_a$, we obtain
\begin{align}
\begin{aligned}
\int \prod_{a=1}^g d \Lambda_a C_M^\Lambda &=  |\Theta|^2 \left[  
\prod_{I=1}^{M}\mu_I^{m_I} \bar \mu_I^{\bar m_I}  \right] \langle \prod_{I=1}^{M} 
V_{\alpha_I} (z_I) \rangle^{\Sigma_g}
\end{aligned}
\end{align}
%\textit{We are missing the $\delta$ function again}
with
\begin{align}
|\Theta|^2 =   q^{-\sum_\nu (m_\nu + \bar m_\nu + k w_\nu)} e^{\frac{3k}{4}U_g}  \prod_{I < J} | E (z_{I} , z_ J ) |^{k} 
\prod_{I} |\sigma (z_I)| ^{-2 k}\, .
\end{align}
Here we should replace $\mu_I$ by the right hand side of \eqref{mug} (and similarly for $\bar \mu_I$).
By applying this result to the computation of \eqref{Cg} we can successfully establish the central 
result \eqref{corrSL2Rg} of this subsection. Thereby, we have now embedded all correlation functions 
of the symmetric product orbifold, including all higher genus contributions, into the Kac-Wakimoto 
free field theory. From here on, the uplift to string theory works exactly as in the genus $g=0$ 
case, only that the string amplitudes are now obtained from worldsheet correlation functions on a
surface $\Sigma_g$ of genus $g$ and the integrations over worldsheet insertion points are 
accompanied by integrations over surface moduli.

\section{Conclusions and Outlook}
\label{sec:conclusion}

In this work we have managed to rewrite the perturbative expansion of symmetric
product orbifold for a parent theory $\mathcal{M}$ with state space \eqref{eq:parentM}
in terms of string amplitudes of some (non-geometric) bosonic string theory that
is defined through the action \eqref{eq:S}. The precise relation between CFT
correlation functions and string theory amplitudes is given in eq.\
\eqref{amplSL2R} for the leading tree level contributions. The bosonic string
theory is based on a free field theory with Kac-Wakimoto field content, i.e. it
contains one $\beta \gamma$ system with central charge $c=2$ and a linear dilaton
$\phi$ with background charge $Q_\phi = b = 1/\sqrt{k-2}$. Hence, the total central
charge is indeed $c=26$, as required by the no-ghost theorem of Brower, Goddard and
Thorn. The theory \eqref{eq:S} contains two interaction terms that do not break the
current algebra symmetry of the free field theory. We denote these interaction
terms by $S^\pm$ where $\pm$ refers to the sign of the spectral flow automorphism
that is used to build the interaction terms. The interaction term $S^-$ is somewhat
conventional and a lose relative of the screening charge of second kind that was
first found by Bershadski and Ooguri. In our context its insertions create poles
of the covering map $\Gamma$ or, equivalently, sheets of the branched coverings.
If the second interaction terms was absent, each insertion of $S^-$ would reduce
the genus of the covering by one until one reached the minimal genus $g=0$. But
when the second interaction term $S^+$ is added something more interesting can
happen. The insertion of $S^-$ can then be thought of as creating two movable
holes that may each trap a pair of interactions $S^+$. Such trapped pairs turn
out to be dual to switching on the marginal deformation with the operator
\eqref{marginalop} in the symmetric product orbifold. This is a very remarkable
mechanism that is based on some very unusual properties of the second screening
charge $S^+$ in our string theory \eqref{eq:S}. The proof of the relation
\eqref{amplSL2R} between string theory amplitudes and orbifold correlators
was based on two central ingredients. On the one hand, it made important use
of the fact that the string theory promotes the covering map $\Gamma$ to a
field $\gamma$ on the worldsheet of the string. Through the localization
mechanisms uncovered in \cite{Eberhardt:2019ywk}, this allowed us to relate string 
amplitudes with (unintegrated) correlation functions in a WZNW model. The second key 
ingredient is our new formula \eqref{corrSL2R} that embeds orbifold correlators and 
their deformation into the WZNW model. The simplicity of this embedding is remarkable.
Most importantly it turned out to trivialize the complicated prefactor on the
right hand side of the eqs.\ \eqref{deformedcorr} and \eqref{deformedcorrg}.
\smallskip

{
Let us stress one more time that the deformed symmetric product orbifold theory is 
a Liouville-like theory. In particular, its correlation functions are expected to 
be meromorphic in the momentum parameters $\alpha$ and possess poles at two series
of poles. The first series is located at the solutions of equation 
\eqref{eq:alphaconstraint}. The residues at these poles are computed through free 
field theory calculations involving insertions of the marginal operator 
\eqref{marginalop} that appear in the action of the deformed symmetric product 
orbifold CFT. As we pointed out, the correlation functions must possess a second 
series of poles whose residues can be computed from insertions of the dual 
marginal operator \eqref{dmarginalop}. The analogy with Liouville theory 
suggests that, along with some analytically assumptions for the dependence on 
the central charge, the residues of the two series of poles completely 
determine the correlation functions of the interacting CFT. This explains  
why it is sufficient to focus on the comparison of the relevant free field 
theory computations with an integer number of insertions of the marginal 
operators. While most of the explicit formulas were spelled out for the 
marginal operator \eqref{marginalop}, we have stressed that much of our 
analysis and in particular the central embedding formula \eqref{corrSL2R}
extends to the second series of poles that is associated with the dual 
marginal operator \eqref{dmarginalop}. After the uplift to the Kac-Wakimoto
representation, the poles of the deformed symmetric product orbifold theory 
were seen to form two series that are located in the very same points in the 
space of momenta $j$ at which they are located in the usual WZNW model. For 
the latter, correlation functions are known throughout momentum space, i.e. 
beyond the poles. Of course, we expect the string theory we describe in 
section 5 to coincide with the usual string theory on $AdS_3$ with $k$ units 
of NSNS flux. In order to fully establish this claim one would 
need to compare all residues. While this has not been done yet, we consider 
the tests in \cite{Eberhardt:2021vsx} as strong support. Clearly, it would 
be interesting to extend these checks. It should be possible to develop a 
more general argument to show that all the residues coincide and thereby to 
show that the string theory we introduced in section 5 coincides with 
string theory in $AdS_3$ with pure NSNS flux.}  

There are a number of very interesting extensions of our work that would be
interesting to work out. The first one is to include boundaries and interfaces
in the symmetric product orbifold. In this string theory this amounts to
studying branes and open strings. Building on previous work, see in particular
\cite{Gaberdiel:2021kkp,Belin:2021nck} (and also \cite{Martinec:2022ofs}) it 
should be possible to derive a AdS/BCFT correspondence \cite{Takayanagi:2011zk}. 
For previous works on branes and open strings on AdS$_3$, see, e.g. 
\cite{Bachas:2000fr,Lee:2001gh,Ponsot:2001gt}. Another interesting direction
would be to study the duality relation in the presence of $\mathcal{N}=2$
supersymmetry using the NSR-formalism for superstrings. In this case one
should be able to make contact with a proposal by \cite{Balthazar:2021xeh,
Martinec:2021vpk}. Still within the context of superstrings it would also
seem worthwhile to consider $\text{AdS}_3 \times \text{S}^3$ and to compare
with a symmetric orbifold $\mathcal{M}^N/S_N$, where $\mathcal{M}$ includes
$\mathcal{N}=4$ super Liouville theory as in \cite{Eberhardt:2019qcl}. Last
but not least, one can also study holography for symmetric product orbifolds
on surfaces of higher genus, and in particular on the torus with genus $g=1$
see e.g.\ \cite{Eberhardt:2020bgq} and further references therein. 
\smallskip

Given that we have identified a worksheet model that is dual to the perturbative
expansion of the symmetric product orbifold, it would be very interesting to
explore integrability of this theory. Following standard lore, the screening
charges $S^\pm$ we introduces in writing down \eqref{eq:S} are expected to
generate an interesting quantum algebra that one expects to control the
integrable structure of the worldsheet model. It would be very interesting
to exploit such structures to write down e.g. TBA-like equations that would
allow to compute the anomalous dimensions of operators in the symmetric
product orbifold along the line of CFTs that is generated by the marginal
operator \eqref{marginalop}. Given the somewhat unusual properties of our
screening charge $S^+$, this direction seems particularly interesting. If
successful, it could also make contact with a regime in which our string
theory admits a sigma model description as a string theory on $AdS_3$.

Finally, it would certainly be extremely interesting to extend this approach
to the engineering of perturbative string theory duals to higher dimensional
CFTs, and in particular to ${\mathcal N}=4$ supersymmetric Yang-Mills theory 
for which some preliminary explorations can be found in \cite{Gaberdiel:2021qbb,
Gaberdiel:2021jrv}.\footnote{Similar ideas were also formulated in 
\cite{Aisaka:2011up}.} 

\bigskip

\noindent
\textbf{Acknowledgements:} We wish to thank Federico Ambrosino, Till Bargheer, Lorentz Eberhardt, Bob Knighton, 
Sean Seed, Alessandro Sfondrini, Vit Sriprachyakul, Yu-ki Suzuki, Joerg Teschner, Takashi Tsuda and Edward Witten 
for interesting discussions and comments. Thanks also to Bob Knighton, Sean Seet and Vit Sriprachyakul for 
sharing a preliminary version of their upcoming paper \cite{Knighton:2023mhq}. 
The work of Y. H. is supported by JSPS Grant-in-Aid for Scientific Research (B) No.\ 23H01170 and 
JSPS Grant-in-Aid for Transformative Research Areas (A) No.\ 21H05187. This project also received funding from 
the German Research Foundation DFG under Germany's Excellence Strategy - EXC 2121 Quantum Universe - 39083330.

%\bibliographystyle{JHEP}
%\bibliography{AdSCFT}

\providecommand{\href}[2]{#2}\begingroup\raggedright\endgroup

\end{document}